\documentclass[11pt,preprint]{aastex}
\usepackage{psfig}

\newcommand{\beq}	{\begin{equation}}
\newcommand{\eeq}	{\end{equation}}
\newcommand{\beqa}	{\begin{eqnarray}}
\newcommand{\eeqa}	{\end{eqnarray}}

\newcommand{\knut}	{\tilde\kappa_\nu}
\newcommand{\krho}	{{k_\rho}}
\newcommand{\kt}	{{k_T}}

\newcommand{\lt}	{\tilde L}
\newcommand{\nut}	{\tilde \nu}
\newcommand{\nch}	{{\nu_{\rm ch}}}
\newcommand{\rch}	{R_{\rm ch}}
\newcommand{\rct}	{{\tilde R_c}}
\newcommand{\rt}	{{\tilde r}}
\newcommand{\rmt}	{{\tilde r_m}}
\newcommand{\rmti}	{{\tilde r_{m,\,\rm int}}}
\newcommand{\rdd}	{{R_{\rm dd}}}
\newcommand{\rddt}	{{\tilde R_{\rm dd}}}
\newcommand{\tch}	{T_{\rm ch}}

\begin{document}


\lefthead{Chakrabarti & McKee}
\righthead{Far-Infrared Spectral Energy Distributions of Protostars and Dusty Galaxies: I}

\title{Far-Infrared Spectral Energy Distributions of Embedded Protostars and Dusty Galaxies: I. Theory for Spherical Sources}

\author{
Sukanya Chakrabarti\altaffilmark{1} and
Christopher F. McKee\altaffilmark{1,2}}

\altaffiltext{1}{
Department of Physics, University of California at Berkeley, Mail Code
7300, Berkeley, CA 94720 USA; sukanya@astro.berkeley.edu.}

\altaffiltext{2}{
Department of Astronomy, University of California at Berkeley, Mail Code
3411, Berkeley, CA 94720 USA; cmckee@astro.berkeley.edu.}

\begin{abstract}
We present analytic radiative transfer solutions for the spectra of
unresolved, spherically symmetric, centrally heated, dusty sources.
We find that the dust thermal spectrum possesses scaling relations
that provide a natural classification for a broad range of sources,
from low-mass protostars to dusty galaxies.  In particular, we find
that, given our assumptions, spectral energy distributions (SEDs) can
be characterized by two distance-independent parameters, the
luminosity-to-mass ratio, $L/M$, and the surface density, $\Sigma$,
for a set of two functions, namely, the density profile and the
opacity curve.  The goal is to use SEDs as a diagnostic tool in
inferring the large-scale physical conditions in protostellar and
extragalactic sources, and ultimately, evolutionary parameters.
Our approach obviates the need to use SED templates in
the millimeter to far-infrared region of the spectrum; this is a
common practice in the extragalactic community that relies on observed
correlations established at low redshift that may not necessarily
extend to high redshift.  Further, we demarcate the limited region of
parameter space in which density profiles can be inferred from the
SED, which is of particular import in the protostellar community as
competing theories of star formation are characterized by different
density profiles.  The functionality of our model is unique in that in
provides for a self-consistent analytic solution that we have
validated by comparison with a well-tested radiative transfer code
(DUSTY) to
find excellent agreement with numerical results over a parameter space
that spans low-mass protostars to ultra-luminous infrared 
galaxies (ULIRGS).   
\end{abstract}

\keywords{galaxies: formation---galaxies: starburst---infrared:
galaxies---radiative transfer---stars: formation}


\section{Introduction}

Stars and galaxies are born in dusty environments, shielded from
view in the optical and often in the near-infrared. Radiation emitted 
by protostars and newly formed stars is absorbed by dust and 
re-radiated at infrared and longer wavelengths. One of the primary
tools for determining the physical parameters of these sources 
is the spectral energy distribution, or SED.

	The classic example of the use of SEDs to infer the nature
of the underlying source is the classification of low-mass
protostars based on the slope of their near-mid IR
spectra (Lada \& Wilking 1984; Lada 1987;
Adams, Lada, \& Shu 1987):
Class I objects, with $d(\nu F_\nu)/d\ln\nu<0$ 
in the 2-20 \micron\ region of the spectrum, are identified
with protostars; Class II objects, with $0<d(\nu F_\nu)/d\ln\nu
\la 2$, are identified with classical T Tauri stars; 
and Class III objects, with $2\la d(\nu F_\nu)/d\ln\nu\leq 3$,
are reddened stars approaching the main sequence.
Subsequently, a fourth category,
Class 0, was added for sources that are so embedded that
it is generally not possible to measure the slope of 
the SED in the near-mid IR (Andre, Ward-Thompson, \& Barsony
1993).

The physical interpretation of observations of
massive protostars 
has proven to be a more challenging task.  
The task of interpretation is complicated by
the greater extinction in high-mass star-forming
regions and by the fact that massive stars often
form in clusters. As a result,
there is little
consensus on either the evolutionary parameters or the source
parameters characteristic of massive star-forming regions.  
Even such a fundamental parameter
as the formation time of a massive star has been uncertain by orders
of magnitude.  Nakano et al. (2000) interpreted 
near-IR spectroscopic data of a source in Orion as indicating
a formation timescale of the order of $10^{3}$ yr.
Osorio, Lizano, \& D'Alessio (1999, henceforth OLD99) estimated
formation timescales of order $10^4$~yr from modeling the SEDs
of hot cores.  Estimates 
of formation timescales based on extrapolating the
theory of low-mass star formation
give formation times greater than $10^{6}$~yr, which is a significant
fraction of the lifetime of the main-sequence lifetime of a massive
star (McLaughlin \& Pudritz 1997, Stahler, Palla, \& Ho 2000).  
Subsequently, McKee
\& Tan (2002; 2003, henceforth, MT03) 
developed the Turbulent Core Model
for high-mass star formation, which incorporates the effects of
the supersonic turbulence and high pressures observed in massive
star-forming regions (Plume et al. 1997) and predicts formation
timescales of the order of $10^{5}$~yr.  

	Other basic characteristics of high-mass star-forming regions,
also remain uncertain.
Only recently has it become possible to observe
individual massive star-forming cores
(Beuther \& Schilke 2004, Cesaroni et al 1999, Fontani et al 2004), 
yet their properties remain uncertain because of 
the limited information on the SED.
The density profile in high-mass star-forming regions is
known only to be 
in the range of 1 to 2, with
significant error bars 
(Shirley et al 2002, Beuther et al 2002, Jorgensen et al 2002, van der
Tak et al 2000, Mueller et al 2002).  On the other hand,
OLD99 have argued for a more precise value,
stating that their data on
dust continuum emission from hot cores are best fit by
so-called logatropic
density profiles, with a power law index of 1.
A self-consistent, analytic methodology for the inference of source
and evolutionary parameters from the observed far-IR SEDs of
protostellar regions would allow us to derive the large scale physical
conditions of star-forming regions, and ultimately to discriminate
between competing theories of star formation.

	A significant fraction of the star formation in the
universe comes from dusty galaxies (Genzel \& Cesarsky 2000), 
including ultra-luminous infrared galaxies (ULIRGs), with
luminosities between $8\;\mu$m and 1~mm in excess of
$10^{12}L_\odot$ (Soifer et al. 1984), and sub-millimeter
galaxies at high redshifts (Smail et al 1997, Barger et al 1999a, 
Blain et al 2002),
which appear to be even more luminous.
SEDs are a primary tool for inferring the properties of
these sources. The lack of data on high-redshift sources
has led to the 
use of SED templates from low-redshift sources to infer
star formation rates in high-redshift sources (e.g., Xu et al
2001, Carilli \& Yun 1999, Yun \& Carilli 2002, henceforth YC02),
but it is not known whether these templates are valid.
The discovery of massive, luminous sub-millimeter
galaxies may warrant a revision of the star formation
history of galaxies at high redshift (Blain et al. 2002) and 
has been suggested as being
problematic for hierarchical formation scenarios (Genzel et al.
2003). The importance of understanding the SEDs of 
high-redshift galaxies is highlighted by the work of
Chapman et al. (2004), who have proposed that higher temperatures
in these sources mean
that current sub-mm surveys may have missed more than half
of the most luminous, dusty galaxies at $z\sim 2$.

	The simplest model for an SED is to assume that it
is due to an isothermal distribution of optically thin dust 
(Hildebrand 1983).  The next level of sophistication is to allow for
the dust to be optically thick above some critical frequency, while
still considering the entire dust envelope as being characterized
by a single temperature, e.g., Yun \& Carilli 2002, henceforth YC02.  
A difficulty with such single-temperature models is that they
often require $\beta$ to be less than the value
appropriate for uncoagulated grain models ($\beta\simeq 2$;
Weingartner \& Draine 2001). This difficulty can be overcome
if the spectrum is assumed to result from the superposition
of two blackbodies at different temperatures
(Dunne \& Eales 2001).  

	In reality, the dust temperature varies continuously.
Variation of the dust temperature has been taken into account
in work on low-mass star formation beginning with the pioneering
work of Larson (1969), who introduced a physically motivated
approximation for the temperature profile.
Adams \& Shu (1985) presented an approximate numerical
radiative transfer model based on this
form of the temperature profile and showed that
they could approximately satisfy radiative equilibrium. 
With this model, they
inferred stellar masses and accretion rates for
their favored collapse model, the singular isothermal sphere.  
This paper heralded the 
beginning of a stream of papers on more refined numerical modeling of 
low-mass protostellar spectra, with inclusion of effects for more evolved sources.
Kenyon, Calvet, \& Hartmann (1993) developed an approximate
numerical approach in which the temperature profile is
calculated from the radiative diffusion equation in
the optically thick part of the envelope and from
radiative equilibrium in the optically thin part.
OLD99 adopted this approach in their calculation
of the SEDs of massive protostars.
Numerical modeling has also enabled the modeling of axisymmetric
sources, including disks (Efstathiou \& Rowan-Robinson 1991;
Kenyon et al 1993; Whitney et al 2003).
However, analytic radiative transfer models for even the simplest
spherically symmetric systems remain rare.  An
exception is that of Adams (1991), who presented an analytic solution for the
specific intensity of protostellar cores 
at millimeter and sub-millimeter wavelengths; however, he 
did not evaluate the accuracy of his method.

	While numerical models permit one to calculate the
SED of a given source with exquisite accuracy, what is 
lacking is any general understanding of how the SED depends
on the underlying source parameters. 
A significant step in addressing this problem
was taken by Ivezic \& Elitzur (1997,
henceforth IE97), who emphasized the importance of
scaling relations in determining the spectrum of dusty
sources. They showed that the spectrum of a spherically
symmetric, dusty source is determined by four parameters:
the dust destruction radius, $\rdd$, which depends on
the luminosity of the source (see Appendix A); the thickness
of the shell around the source, $y\equiv R_c/\rdd$, where
$R_c$ is the radius of the cloud in which the source is
embedded; the optical depth through the shell at some
frequency; and the temperature of the central source. In addition,
the spectrum depends on two functions, the density distribution,
$\rho(r)$, and the opacity, $\kappa_\nu$. Using these
scaling properties of dust
emission, they developed a numerical code, DUSTY, 
that is very useful in inferring the physical conditions
in dusty sources.

	Our objective in this paper is to develop
an analytic theory for the SED of spherically symmetric,
dusty sources. We assume that the dusty envelope surrounding
the central source of radiation is sufficiently opaque that
the resultant SED is approximately independent of
the temperature of the central source and of the properties
of the dust destruction front.  Within our parameter space,
$100 \rm g~cm^{-2}\ga \Sigma \ga 0.03 \rm g~cm^{-2}$,
the optical depth is large enough that essentially
all the stellar radiation is absorbed by dust and re-emitted.
Because of the large optical depth, the near-infrared and
optical spectrum, which does depend on the source spectrum,
is heavily attenuated.  For a given form for the 
density distribution (e.g., a power-law) and the opacity
(e.g., that of Weingartner \& Draine 2001), the emergent
spectrum depends on three parameters: the luminosity, $L$;
the dust mass, $M_{\rm dust}=M/Z_{\rm dust}$,
where $Z_{\rm dust}$ is the mass fraction of dust; 
and the radius of the source,
$R_c$. The {\it shape} of the SED is independent of the 
distance, and therefore depends on only two parameters, which
we take to be the luminosity-to-masss ratio, $L/M$, and
the surface density of the source, $\Sigma\equiv M/(\pi R_c^2)$.
As we shall show, it is possible to infer these two
distance-independent parameters, {\it and therefore the
complete shape of the SED}, from just two colors
(provided, of course, that the assumptions underlying
our model are correct).
Since the pressure in a self-gravitating gas is
of order $G\Sigma^2$, determination of the surface density
allows one to infer the pressure in the source.
If the distance is also known, we can
infer the luminosity, dust mass, and physical size
of a protostellar region--even if the source is unresolved. 
However, it is generally not feasible to infer the density
profile from the far-IR SED of an unresolved source; as we shall see later, this is feasible for extended envelopes, which
are large compared to the radius of the effective
photosphere, and for envelopes that emit most of their radiation at wavelengths shorter than 30$\mu\rm m$.

We find that the spectra are characterized by three
frequency regimes: Low frequencies, which are always in
the Rayleigh-Jeans limit [$h\nu<kT(R_c)$];
intermediate frequencies, which are not necessarily in
the Rayleigh-Jeans limit, but are low enough that the envelope
is transparent; and high frequencies, where the spectrum
is determined by competition between opacity effects and
the Wien cutoff. Having
found formal solutions in these three
regimes, we present a joint solution
within a conceptually simple and physically motivated framework
in which the emission in each frequency regime comes from a
shell of some thickness, centered at some radius, and is attenuated
by the intervening optical depth.  We adopt a self-consistent
temperature profile that characterizes the emission coming
from the vicinity of an effective Rosseland photosphere.  
The slope of this profile is set by the condition that the
emergent luminosity equal the input luminosity.  
We have tested the
accuracy of our analytic solution by making detailed comparisons with
DUSTY.

The organization of the paper is as follows: in \S 2 we outline the
general formulation of the problem, and specify the forms of the
density profiles that we consider as well as the dust opacity.  In  \S
3 we introduce the characteristic parameters that are an integral part
of our formalism.  In \S 4 we present the formal solution to the
problem in the three frequency regimes that are characteristic of the
dust thermal spectrum, and in \S5 we present the joint solution for
the SED.  In \S 6 we discuss luminosity conservation and the resultant form
of the temperature profile.  \S 7 presents the accuracy and range of
applicability of our solution.  \S 8 is dedicated to an explanation of
our results and analysis, with particular attention to the shape of
the SED and its dependence on two parameters, the feasibility of
inferring density profiles from the SED, and the emergent
three-component spectrum for highly extended envelopes.  
In Paper II, we shall present the far-IR SEDs and inferred source
and evolutionary parameters for a broad range of sources, from low
mass protostars, to massive protostars and ULIRGS.

\section{Formulation of the Problem}

We consider a centrally concentrated source of radiation surrounded by
a homogenous, spherical distribution of dust.  Consideration of a
central source of radiation, while appropriate for protostellar
sources, is approximately valid for ULIRGs, particularly if the
far-IR emission is predominantly powered by an extended starburst.  As
such, our method is more applicable for ULIRGs largely powered by
dust-enshrouded AGN or compact star clusters, and for super-star clusters
in starburst galaxies and ULIRGs.  
A relevant finding here is that of Soifer (et al 2000) -
they find that that a large fraction of the mid-infrared emission stems from very compact systems.  
If the source is unresolved, our method is
applicable when the dust temperature in the beam is dominated by the central source.  
If there are multiple sources present, or if the background temperature is
significant, then the angular resolution needs to be sufficient to meet the above condition.  If the angular resolution of the beam is sufficient to resolve the source, our method can be applied to compute the emergent SED given that all the flux from the source is included in the beam. 
Our assumption of spherical symmetry limits us to consideration of young protostellar sources
in which any accretion disk is small enough with respect to the
surrounding envelope that it does not significantly affect the SED.
 We approximate the density distribution by a power law in
radius. We also assume that the dust is homogeneous; the
effect of clumping will be considered in a future paper (Chakrabarti \& McKee, Paper III).

We neglect scattering of radiation by dust grains and consider only
the thermal emission in computing the emergent SED, as the scattering
efficiency scales as $\lambda^{-4}$ and is thus important only at very
short wavelengths.  We consider dust shells sufficiently opaque that
the dust destruction front is highly obscured, and thus does not
significantly affect the far-IR spectrum (see Appendix A).  This also
implies that the observed SED is independent of the spectrum of the
central source of radiation.  We consider the opacity to be a function
of frequency only; we do not consider the temperature
dependence of the opacity that can be caused by evaporation of grains.  
Although we do not treat the temperature dependence of the
opacity explicitly, we give a simple prescription to implement it in
Paper II.  These assumptions are discussed in more detail in \S 7.

We find that we get excellent agreement with the numerical solution
from DUSTY for the far-IR SED with the adoption of a power law
temperature profile.  We have specified the variation of the slope of
the temperature profile over a parameter space that spans a range of
optical depths of a factor of $\sim 1000$, through a combination of
heuristic arguments and numerical calibration with DUSTY.  Thus,
although the actual temperature profile is not a pure power law, the
mm to far-IR emission can be well described by an effective power-law
temperature profile that characterizes the emission coming from close
to the Rosseland photosphere.  Most of the observed emission
originates outside the Rosseland photosphere, since emission at high
frequencies is attenuated by the intervening optical depth, and
emission at low frequencies comes from outside the $\tau=1$ surface at
the peak of the SED.  

As explained in \S 6, we adopt a temperature profile that is a result
of imposing the self-consistency criterion that the input luminosity
be equal to the emergent luminosity.  We find that, for a given
density profile and dust model, the slope of the temperature profile
is a function of one dimensionless parameter, insofar as most of the
emitted flux is longwards of $\lambda\sim$ 30 $\mu$m, where the
opacity is approximately 
a power law in frequency.  

\subsection{Density Profile} 

Observations of low 
and high mass star forming regions (Shirley et al 2000, Shirley et al
2002, Beuther et al 2002, Jorgensen et al 2000, van der Tak et al
2000, Mueller et al 2002) have found power law density profiles,
$\rho(r)\propto r^{-k_{\rho}}$, with density power law index,
$k_{\rho}$, in the range of 1 to 2.  As discussed in more detail in
\S 8, within our formalism we approximate $k_{\rho}=1$ with $k_{\rho}=1.1$,
as the emitted spectra agree nearly exactly.  We note that most of these
authors have found power law density profiles from a combination of
fitting to the observed SED and intensity profiles.  (We defer detailed
discussion of the feasibility of inferring density profiles from the
SED alone to \S 8.)
The observed prevalence of power law density
profiles in protostellar envelopes can be heuristically understood in
the context of a self-similar, turbulent medium (MT03).  
The self-similarity should be broken only on small
scales, by thermal motions (MT03), as they are observed
to do in regions of low-mass star formation (Blitz \& Williams 1997).
However, such a picture of a time-stationary,
spherical, self-gravitating turbulent structure is necessarily highly
approximate, and large fluctuations, both spatial and temporal, are to 
be expected. We defer consideration of these fluctuations, which
correspond to clumpiness, to a future paper.

	It is important to note
that the observations of massive star forming regions 
cited above probed scales of {\it clumps},
which are massive enough to form a cluster of
stars, rather than of {\it cores}, which will form
a single star or binary.  (We follow the clump-core terminology
introduced in Williams, Blitz \& McKee 2000.) 
However, as we shall show in Paper II,
the photosphere of a high-mass protostar is within
its natal core. Cores are too small to
have measured density profiles yet.
We assume that they also have power-law density profiles,
which is consistent with MT03's assumption
that clumps and cores are part of a
self-similar structure.
We assume that the cores have a definite
outer radius at $R_c$, 
beyond which the density drops rapidly; we assume that the emission
from beyond $R_c$ is negligible.

While the simplest SED models of star-forming galaxies do not consider
a density variation (e.g. YC02), more sophisticated radiative transfer
models (Efstathiou et al 2000) have considered starburst galaxies as
an ensemble of optically thick clouds heated by the newly formed
stars.  The assumption of power-law density profiles for ULIRGs is
a very approximate representation of the complex morphology of merging
systems; it may be a good approximation for the clouds that comprise
the ULIRGs, however.

\subsection{Dust Opacity}

The results presented in this paper are based on the Weingartner \&
Draine (2001) (henceforth WD01) $R_{V}=5.5$ dust model.   WD01 is an
extension of the original Draine \& Lee (1984) dust model, with a size
distribution developed to reproduce the observed extinction curve for
a variety of environments, parametrized by the ratio of the visual
extinction to reddening, $R_{V}$.  For the diffuse ISM, $R_{V}$ is
observed to be approximately 3.  Higher values have been observed for
dense clouds (Strafella et al 2001, Kandori et al 2003, Vrba et al
1993) and star-forming galaxies (Calzetti 2000).  

The WD01 model has a simple composition, consisting of carbonaceous
grains and silicates.  WD01's $R_{V}=5.5$ model has a substantial depletion of
the smallest carbanaceous grains relative to the $R_{V}=3.1$ model,
leading to a difference in the opacity curves in the near-IR.
However, WD01's best fit models for $R_{V}=5.5$ and $R_{V}$=3.1 have
no difference in the far-IR extinction curves, and thus imply no
substantial grain growth for sizes on the large end of the size
distribution, i.e., $a\sim 0.1 \mu$m.  As we later show, for densities
typical of protostellar regions and dusty galaxies, it is unlikely
that significant coagulation occurs within a few free-fall times to
affect the spectrum in the mm to far-IR region of the spectrum.  For
uncoagulated grain models such as WD01, the mm to far-IR variation of
the opacity with frequency (in the range of 3 $\rm mm$ to 30 $\mu \rm
m$) is well represented as a power 
law with slope, $\beta=2$.

	The opacity normalization per gram of dust,
$\kappa_{\nu_{0}}$, depends on the metallicity.
Let $Z_{d}$ be the mass fraction of
dust, equal to 1/105.1 for solar abundances (WD01)
and $\delta$ be the dust-to-gas ratio relative to solar.  
At a fiducial wavelength of
$\lambda_{0}=100 \mu\rm m$, 
WD01's opacity is
$\kappa_{\nu_0}=0.27\delta $,
independent of $R_V$.
The WD01 models reproduce the observed extinction curves for the Milky
Way, 
as well as regions of low metallicity, such as the LMC and the SMC.  
Since WD01 is the simplest grain model
that is able to reproduce the
observed observed extinction over a wide range of metallicities,
we have adopted it 
as our fiducial dust model.

	The principal omission in the WD01 model is the lack
of ice mantles. Spectroscopic studies have identified
ice spectral features in protostellar environments and in dusty
galaxies (Allamandola et al 1992, Tielens et al 1984, Spoon et al
2002).  Based on observations with the Submillimeter Wave Astronomy
Satellite, Bergin et al. (2000) have argued that most of the
oxygen in star-forming regions
is frozen out onto the dust grains in the form of molecular
ices.
Preibisch et al. (1993) 
have modeled dirty ice mantles, and find
significant variations in the opacity, depending on how the mantles
are distributed among the grains and on the volume fraction of
carbonaceous material; a typical increase in the far-infrared 
opacity is a factor $\sim 2$.
Pollack et al (1994) considered 
detailed grain compositions, including the contribution from water
ice. If most of the oxygen is in the form of water ice, they
find that $\kappa_{\nu_0}\simeq 1$ cm$^2$ g$^{-1}$ at 100 \micron;
however, at $=\lambda_0=1$~mm, the opacity for spherical grains
of average radius 1 \micron\ is $4.2\times 10^{-3}$ cm$^2$ g$^{-1}$,
not  that different from WD01 ($\kappa_{\nu_0}=3.0\times 10^{-3}$
cm$62$ g$^{-1}$).  However, the steep power law index for the opacity, $\beta=2.7$, 
found by Pollack et al. does not appear to be consistent with observations.  
Pollack et al. show that for typical conditions
in regions of high-mass star formation ($n_{\rm H}\sim 10^6$
cm$^{-3}$), ice sublimates at $T\simeq 110$~K.
We conclude that sigificant uncertainties remain in
the effect of ice mantles on the far-infrared opacity.  We show later in \S 3 that the effective 
increase in opacity normalization that ice mantles produce may be simply treated within our formalism.

	A further complication to consider is the possible
growth of grains due to coagulation (Pollack et al. 1994;
Ossenkopf \& Henning 1994).
The time scale for coagulation depends on the mean
relative speed of the grains, which is likely due by 
turbulence (Draine 1985; Lazarian \& Yan 2002) 
in dense cores.
The outcome of a collision between grains depends on whether or not the
grains stick together, i.e., if the surface potential energy is
comparable to or larger than the kinetic energy.  
Recent numerical experiments (Poppe et al 2000) have shown that
relative velocities of order $1 \rm m/s$ allow sticking probability of
order unity, with a sharp decline at $v\sim 5 \rm m/s$.  
Therefore, the timescale for grain coagulation is approximately
\begin{equation}
t_{\rm coag}\sim \frac{1}{n\sigma v_{rel}}=6\times 10^{6}\rm yr
\left(\frac{10^{5}\rm cm^{-3}}{n_{H}}\right)\left(\frac{a}{0.1\mu\rm m}\right)\left(\frac{1 \rm m~s^{-1}}{v_{\rm rel}}\right)\;,
\end{equation}
where we have considered the approximate case of coagulation of
identical, spherical grains
of size $a$.
To determine whether coagulation will be
effective within timescales
relevant for protostellar evolution, we compare this coagulation time
scale to the free fall time.  For coagulation to affect the spectrum in the 
far-IR range would require grain growth to the Rayleigh limit, 
i.e., to sizes $a \sim \lambda/2\pi,
a \ga 5$~\micron, (to affect emission at $\lambda \ga 30 \micron$).
This gives:
\begin{equation}
\frac{t_{\rm coag}}{t_{\rm ff}}\sim 2000\left(\frac{10^{5}\rm
cm^{-3}}{n_{H}}\right)^{1/2}\left(\frac{a}
{5 \mu \rm m}\right)\left(\frac{1 \rm m~s^{-1}}{v_{rel}}\right)\;.
\end{equation}
Therefore, for typical estimates of the densities, $n_{H}\sim
10^{6}\rm cm^{-3}$, we see that the coagulation timescale is more than 
several hundred times the free-fall time for $a\sim 5\rm \mu\rm m$, which corresponds to $\lambda=30\mu\rm m/2\pi$.  This conclusion is consistent with 
the results of Chokshi et al. (1993), who
found a relatively small increase
in grain size (approximately a factor of 2) in dense cores.
We conclude that coagulation in star-forming regions is unlikely
to significantly alter the far-infrared opacity.

\section{Characteristic Parameters}

	Stars have well-defined photospheres, but dust clouds
do not. Nonetheless, the SED of a dusty source can be described
in terms of characteristic radius, $\rch$, 
and characteristic temperature, $\tch$, such that
\beq
L\equiv 4\pi\lt \rch^{2}\sigma \tch^{4}\; ,
\label{eq:L}
\end{equation}
where $\lt$ is a number of order unity determined below in order
to secure better agreement with the actual SED.
We determine $\rch$ and $\tch$ by requiring that
a characteristic optical depth at frequency $\nch\equiv k\tch/h$ is
unity,
\begin{equation}
\tau_{\rm ch}=\kappa_{\rm ch}
\int_{1}^{{R_c}\rightarrow\infty}\rho(\tilde{r})d\tilde{r}=
\frac{\kappa_{\nch} \rho(\rch)R_{\rm ch}}{k_{\rho}-1}=1\; ;
\label{eq:tauch}
\end{equation}
note this characteristic optical depth ignores the cut off in
the density at the edge of the cloud, which is at $R_c$.
$\rch$ and $\tch$ are therefore the approximate photospheric
radius and temperature; more accurate photospheric values are
given in \S \ref{S:accuracy} below.

	We now express the characteristic parameters in terms of the physical
source parameters, the surface mass density, $\Sigma\equiv M/\pi
R^{2}$, and the luminosity to mass ratio, $L/M$.  
We assume that $\nch$ lies within the frequency regime in
which the opacity is a power law,
\beq
\kappa_{\nu}=\kappa_{\nu_{0}}(\nu/\nu_{0})^{\beta}~~~~~(30~\mu{\rm m
	\la \lambda \la 1~mm}).
\eeq 
Solving equations (\ref{eq:L}) and (\ref{eq:tauch}),
we evaluate the two parameters that we use
to describe our solutions,
\begin{equation}
\rct\equiv\frac{R_{\rm c}}{R_{\rm ch}}=
\left\{\frac{(L/M)\Sigma^{(4+\beta)/\beta}}{4\sigma \tilde{L}}
\left[\frac{(3-k_\rho)\kappa_{\nu_0}}{4(k_{\rho}-1)T_0^\beta}\right]
^{4/\beta}\right\}^{-\frac{\scriptstyle \beta}
{\scriptstyle 2\beta+4(k_{\rho}-1)}}\; ,
\label{eq:rct}
\end{equation}
and 
\begin{equation}
T_{\rm ch}=\left\{\frac{L/M}{4\sigma \tilde{L} 
\Sigma^{\frac{3-k_{\rho}}{k_{\rho}-1}}}
\left[\frac{4(k_{\rho}-1)T_{0}^{\beta}}{(3-k_{\rho})\kappa_{\nu_{0}}}\right]
^{\frac{2}{k_{\rho}-1}} \right\}
^{\frac{\scriptstyle k_{\rho}-1}{\scriptstyle 2\beta+4(k_{\rho}-1)}}\; .
\label{eq:tch}
\end{equation}
In general, $\lt$ is a function of $\rct$, the form of which
is specified in \S 6.  
It is important to note
that while we have defined the characteristic parameters in terms of a
power law opacity for simplicity, our solution for the emergent SED is
valid for an arbitrary opacity curve, 
as we show later.  

	The utility of equations (\ref{eq:rct}) 
and (\ref{eq:tch}) is to allow an analytic translation
between the SED variables, $\rct$ and $\tch$, which govern the shape
of the SED, and the source parameters, $L/M$ and $\Sigma$.  The
luminosity-to-mass ratio and the surface density are independent of
distance, and therefore, the SED variables, $\rct$ and $\tch$, are
also.  Once the SED variables are determined from two observed color
ratios, we may then immediately solve for $L/M$ and $\Sigma$ using
these equations.

	The SED variable $\rct=R_{c}/R_{\rm ch}$ is a dimensionless measure of
the size of the region that the observed far-IR SED probes.  When it is large, the observed emission arises from a wide
range of radii; when it is somewhat larger than unity,
the emission comes from a narrow range. Values of
$\rct$ less than unity are generally not meaningful since
the resulting spectrum is sensitive to our assumption
that the emission at $r>R_c$ drops to zero. As we
shall see below, dense dust shells, such as those characteristic of
massive protostars, starbursts and AGN, have smaller values of $\rct$
than lower density dust shells, such as those characteristic of
low-mass protostellar envelopes.  Thus, for low-mass protostellar
envelopes, we probe the source function over a larger region 
than for high-mass envelopes.

	The angular size of the photosphere is about
\beq
\theta_{\rm ch}\equiv \frac{\rch}{D}\; , 
\eeq
where $D$ is the distance to the source.  
If the total flux, $F\equiv L/4\pi D^2$, is known, then the
angular size is
\beq
\theta_{\rm ch}=\left(\frac{F}{\lt\sigma\tch^4}\right)^{1/2}
\eeq
from equation (\ref{eq:L}). Thus, since $\tch$ determines
the surface brightness of the source, it is possible to
infer the angular size of the source even if it is
unresolved. 
If the distance to the source is also known, we
may then further solve for the 
size of the source, as well as the mass and luminosity.

	In Figure 1a, we have depicted the typical locations of various sources
on the $L/M$ vs $\Sigma$ plot.  Isolated low-mass protostars
appear typically on the lower left corner of this diagram, i.e., at
low surface density, $\Sigma \sim$ 0.05 $\rm g~cm^{-2}$, and low
luminosity to mass ratios, $L/M\sim 1\; L_{\odot}/M_{\odot}$ (Jorgensen
et al 2002). In
clusters, $\Sigma$ is higher since the pressures are higher (MT03).
High-mass protostars, ULIRGs and super-star clusters are typically in regions
of higher surface density, $\Sigma \sim 1 \;\rm g/cm^{2}$ (Plume et al
1997, MT03), and thus appear on the right-hand side of this diagram.
ULIRGs and super-star clusters have comparable $L/M \sim 10\; L_{\odot}/M_{\odot}$ on average (Downes \& Solomon 1998),
(Gilbert \& Graham 2002, Gilbert 2002).  Two sets of lines of constant $\tch$
and $\rct$ are overlaid on this diagram for our fiducial density
profile, $k_{\rho}=3/2$, and our adopted dust model WD01.  
As discussed above, ice-coated grains produce an effective
change in the opacity normalization of a factor of $\sim 2$.  Our
scaling relations (\ref{eq:rct}) and (\ref{eq:tch}) show that this
leads to a small change in the governing SED variables.  The
difference relative to WD01's normalization when ice-coated grains are
considered is depicted in Figure 1b for our fiducial density 
profile.
A more quantitative discussion of the characteristic 
parameters is relegated to \S \ref{S:accuracy}.  
Figures 2a, b, and c present fits to observed data for a low, high-mass protostar and ULIRG, respectively.  Paper II gives a detailed explanation of the application of our methodology to star-forming system, along with SED fits to about a dozen sources, from low-mass protostars to ULIRGs.

\section{Evaluation of $L_{\nu}$}
\label{S:Lnu}

We compute the spectral luminosity as
\begin{equation}
L_{\nu}=4 \pi \int_\rdd^{R_c} j_{\nu} f_{\nu}(r)4 \pi r^{2} dr \; ,
\end{equation}
where $j_{\nu}$ is the emissivity and $f_{\nu}(r)$ is the
escape probability, which is evaluated
in Appendix B.
The inner boundary of the emitting shell is at $\rdd$,
the dust destruction radius, and the outer boundary is at
$R_c$, the radius of the cloud.
At high densities, the dust and gas temperatures are
the same, but at low densities they may differ; in our
equations, $T$ always refers to the dust temperature.
We assume that the dust emissivity is given by the
LTE expression,
$j_{\nu}=\kappa_{\nu}\rho(r)B_{\nu}(T)$.
This approximation breaks down at the dust
destruction front, where the absorption of UV photons
can lead to transient heating of small grains (Draine \& Anderson 1985). However, since we assume
that the cloud is sufficiently opaque that the
dust destruction front is shielded from view, the
LTE approximation should be valid.

In order to evaluate the luminosity, we must specify
the temperature profile. We assume that it can be approximated
as a power law in the vicinity of the photosphere,
\beq
T=\tch\left(\frac{r}{\rch}\right)^{-\kt}=\tch\rt^{-\kt}\; ,
\eeq
where $\tilde r\equiv r/\rch$.
We determine $\kt$ to zeroth-order by imposing the self-consistency condition that the input luminosity equal the emergent luminosity - the procedure is described in detail in \S 6.  Note that this assumed form for the temperature carries
the implicit assumption that the temperature at $\rch$ is $\tch$;
we choose $\lt(\rct)$ choose so as to improve the accuracy
of this approximation. 
Written in dimensionless notation, our expression for the luminosity
then becomes
\begin{equation}
L_{\nu}=4\pi
\rch^{2}4\pi\left(\frac{2h\nu_{ch}^{3}}{c^{2}}\right)\tilde{\kappa}_{\nu}\tilde{\nu}^{3}(k_{\rho}-1)I\; ,
\label{eq:lnu}
\end{equation}
where
\begin{equation}
I=\int_\rdd^{R_c}\frac{f_{\nu}(\tilde{r})\tilde{r}^{2-k_{\rho}}}
{\exp(\tilde{\nu}\tilde{r}^{k_{T}})-1}d\tilde{r}\; ,
\label{eq:I}
\end{equation}
$\nut\equiv \nu/\nch$, $\knut\equiv \kappa_\nu/\kappa_{\nch}$,
and $h\nu/kT(r)=\tilde{\nu}\tilde{r}^{k_{T}}$. 
Note that since the integral is
performed over position, we may take the opacity term, $\kappa_{\nu}$,
outside the integral since we have assumed the opacity is not a
function of position.

	In order to obtain an approximate analytic evaluation
of $I$, we have found it necessary to consider three distinct
frequency regimes, which we denote as low, intermediate, and
high. The low and intermediate frequencies are optically thin.
Low frequencies are in the Rayleigh-Jeans portion of the
spectrum throughout the envelope [$h\nu<kT(R_c)$].
The low-frequency emission
comes predominantly from the outer parts of the shell, as it
is proportional to the mass.  Intermediate
frequencies are in the Wien portion of the spectrum
in the outer envelope, although not near the photosphere.  The
temperature dependence of the intermediate frequency region causes
the higher components of this frequency regime to emanate from
deeper in the envelope, in a sense specified by the temperature variation
of the envelope.  High-frequencies are in the Wien portion of the 
spectrum at the photosphere; the emission originates from a location 
that is due to a tug-of-war between the hotter temperature in
the interior and the intervening optical depth 
that has to be traversed.  

	Once we obtain an approximate expression for $L_\nu$
in each of these frequency regimes, it is necessary to knit
them together into a single expression. To do this, we
have found it convenient to introduce a ``shell'' model,
in which the emission at each frequency comes
from a shell of thickness
$\Delta r_{m}(\nu)$ centered at 
a radius $r_{m}(\nu)$, with a source function
$(2h\nu_{ch}^{3}/c^{2})\exp\left[{-h\nu/kT(\rmt)}\right]$
located at an optical depth 
$\tau_\nu(\tilde{r}_{m})$:
\begin{equation}
L_{\nu}=4\pi
\rch^{2}4\pi\left(\frac{2h\nu_{ch}^{3}}{c^{2}}\right)
\tilde{\kappa}_{\nu}\tilde{\nu}^{3}(k_{\rho}-1)\tilde{r}_{m}^{2-k_{\rho}}
\Delta\tilde{r}_{m}\exp\left[-\frac{h\nu}{kT(\tilde{r}_{m})}-
\tau_\nu(\tilde{r}_{m})\right]\; ,
\label{eq:shell}
\end{equation}
where the optical depth $\tau_\nu$
from $r$ to the surface of the cloud is
\begin{equation}
\tau_{\nu}=\tilde{\kappa}_{\nu}\left(\tilde{r}^{-k_{\rho}+1}-
\rct^{-k_{\rho}+1}\right)  \; .
\label{eq:tau}
\end{equation}
We now proceed to evaluate the parameters of this intuitively
transparent form for the spectral luminosity.

\subsection{Low and Intermediate Frequency Regimes ($\tau_\nu\ll 1$)}

	The 
low and intermediate frequency regimes are characterized by 
$\tau_\nu \ll 1$, so that the escape fraction is approximately unity,
$f_\nu(\rt)\simeq 1$.
Low frequencies are in the Rayleigh-Jeans part of
the spectrum throughout the cloud, so that
$h\nu/kT=\nut \rt^\kt \ll 1$ everywhere and the exponential
in $I$ can be expanded,
\beq
I_{\rm low}=\frac{\rct^{3-\krho-\kt}}{(3-\krho-\kt)\nut},
\label{eq:low}
\eeq
where we have assumed that $\rdd^{3-k_{\rho}-k_{T}}\ll
R_{c}^{3-k_{\rho}-k_{T}}$, so that the dust destruction radius does not
significantly affect the spectrum; equivalently,
the emission is dominated by the outer part of the
envelope.  We note that this
requires $k_{T} < 3-k_{\rho}$.

	Intermediate frequencies are in the Wien part
of the spectrum in the outer part of the cloud
($h\nu/kT(R_c)\gg 1$), and as a result the upper
limit integration can be set to infinity. Since
we have assumed that the dust destruction front
does not affect the spectrum, we then find
\beq
I_{\rm int}=\Gamma\left(\frac{3-k_{\rho}}{k_{T}}\right)
\zeta\left(\frac{3-k_{\rho}}{k_{T}}\right)\frac{1}{k_{T}\tilde{\nu}
^{(3-k_{\rho})/k_{T}}}
\label{eq:int}
\end{equation}
(Gradshteyn \& Ryzhik p. 349).

	Now we combine the formal expressions for the low and
intermediate frequency fluxes into a single expression by observing
the scaling behavior of each.  In this case, we require a harmonic
mean to recover each limiting case, $I_{\rm low-int}^{-1}=I_{\rm
low}^{-1}+I_{\rm int}^{-1}$. The resulting expression for
the luminosity is
\begin{equation}
L_{\nu,\,\rm low-int}\simeq 16\pi^{2}
(\krho-1)\left(\frac{2h\nu_{ch}^{3}}{c^{2}}\right)
\knut\rch^{2}\tilde{\nu}^{3} \left[\frac{\Gamma\zeta}
{(3-k_{\rho}-k_{T})\tilde{\nu}\Gamma\zeta\rct^{-(3-k_{\rho}-k_{T})}
+k_{T}\tilde{\nu}^{(3-k_{\rho})/k_{T}}}\right]\; ,
\label{eq:lowint}
\end{equation}
where the argument $(3-k_{\rho})/k_{T}$ for the Gamma and Zeta 
functions has been suppressed for clarity.  This prescription 
holds for $k_{\rho}$ in the range of 1 to 2 that we are considering.  

	Recall that in our shell model (eq. \ref{eq:shell}), 
we characterize the emission at each frequency
as coming from an optically thin shell centered at a radius
$r_{m}(\nu)$ and having width $\Delta r_{m}(\nu)$.
To apply this equation, we must determine the shell radius and
thickness.
The shell radius $r_{m}$ is also known as the contribution function 
(e.g., Gray p. 278).
As discussed below equation (\ref{eq:low}), most of the 
low-frequency emission comes from the outer part of the 
shell, so we set $\tilde{r}_{m,\,\rm low}=\rct$.  
To determine the shell radius at intermediate frequencies, 
we assume that $h\nu/kT(\rmti)$ is a constant,
\beq
\frac{h\nu}{kT(\rmti)}=\nut\rmti^\kt=C \; ,
\label{eq:C}
\eeq
where $C$ is determined from numerical calibration in \S 7 and \S 8.1.
To recover the limiting cases, the shell radius for
both low and intermediate frequencies can be approximated
by the harmonic mean, so that
\begin{equation}
\tilde{r}_{m,\,\rm low-int}=\frac{\rct
C^{1/\kt}}{\rct\tilde{\nu}^{1/k_{T}}+C^{1/\kt}}\; ,
\end{equation}
As we discuss in detail in \S 8.1, when $\tilde{r}_{m,\,\rm low}=
\tilde{r}_{m,\,\rm int}$, the slope of the emitted spectrum changes 
from the standard Rayleigh-Jeans slope to the flatter intermediate 
frequency slope.  This break frequency, i.e., the frequency at 
which the character of the emission changes from being dominated 
by the cool material on the outside to the emission coming from 
deeper into the envelope in a sense specified by the temperature 
gradient, is then given by:
\begin{equation}
\tilde{\nu}_{\rm break}=C\rct^{-k_{T}}\; .
\label{eq:nubreak}
\end{equation}   
In terms of the temperature at the outer edge of the envelope, the
break frequency is:
\beq
\nu_{\rm break}=\frac{kT(R_c)}{h} \; .
\eeq
 
     In order to recover equation 
(\ref{eq:lowint}) for the luminosity at low
and intermediate frequencies, we must set
the thickness of this emitting region to be 
\begin{equation}
\Delta \tilde{r}_{m}=\left[\frac{\Gamma\zeta e^{h\nu/kT
(\tilde{r}_{m,\,\rm low-int})}}{(3-k_{\rho}-k_{T})\tilde{\nu}
\Gamma\zeta\rct^{-(3-k_{\rho}-k_{T})}+k_{T}\tilde{\nu}^
{(3-k_{\rho})/k_{T}}}\right]\frac{1}{\tilde{r}_{m,\, \rm
low-int}^{2-k_{\rho}}}\; .
\end{equation} 
  
\subsection{High-Frequency Regime}

In the high-frequency regime, the optical depth can be larger than
unity and it is necessary to develop an escape fraction appropriate for
spherical geometry.  In Appendix B we show that an appropriate escape 
fraction for spherical geometry is:
\begin{equation}
f(\tilde{r})\simeq\frac{e^{-\tau_{\nu}}}{1+2\tilde{\kappa}_{\nu}
\left(\frac{k_{\rho}-1}{k_{\rho}+1}\right)
\left(\tilde{r}^{1-k_{\rho}}-\tilde{r}^{2}
\rct^{-k_{\rho}-1}\right)}\; ,
\end{equation}
where $\tau_{\nu}$ is the radial optical depth and is given in
equation (\ref{eq:tau}).  
We note that for a given dust model and density
profile, the optical depth depends on $\tch$ in a scale-free manner so long
as we consider wavelengths longwards of 30~$\mu$m,
where $\kappa_\nu\propto \nu^p$ is a power-law in frequency.

We compute the high-frequency luminosity by the method of steepest
descent.  This method is particularly useful when the integrand is
composed of a rapidly varying function and a slowly varying function,
such as an exponential term multiplied by an algebraic term, which is
indeed the structure of the high-frequency integrand.  
The method of steepest descent approximates
the integral as the algebraic term evaluated at the local maximum of
the exponential and a gaussian centered at the local maximum, having a
width that is proportional to the second derivative of the argument of
the exponential.  The narrowness of this gaussian is an indication of
how local the contribution to the integrand really is.  The narrower 
the gaussian, the better the approximation.  

Inserting our expression for
the escape fraction into equations (\ref{eq:lnu}) and (\ref{eq:I}) 
we find
\begin{equation}
L_{\nu}=16 \pi^2 (k_{\rho}-1)\rch^{2}\tilde{\kappa}_{\nu}\tilde{\nu}^{3}
\left(\frac{2h\nu_{ch}^{3}}{c^{2}}\right)
\int_0^\infty \frac{\tilde{r}^{2-k_{\rho}}
e^{-h(\rt)}
}
{1+2\tilde{\kappa}_{\nu}\left(\frac{k_{\rho}-1}
{k_{\rho}+1}\right)\left(\tilde{r}^{1-k_{\rho}}-\tilde{r}^{2}
\rct^{-k_{\rho}-1}\right)} d\tilde{r}\; .
\label{eq:steep}
\end{equation}
where the argument of the exponential term is
\begin{equation}
h(\tilde{r})=\tilde{\nu}\tilde{r}^{k_{T}}+\tilde{\kappa}_{\nu}
\left(\tilde{r}^{1-k_{\rho}}-\rct^{1-k_{\rho}}\right)\; .
\end{equation}

	The location of the maximum of the exponential term 
[i.e., the minimum of $h(\rt)$] gives the shell radius
at high frequencies according to the method of steepest descent,
\begin{equation} 
\tilde{r}_{m,\,\rm high,\;steep}=\left[\frac{\tilde\kappa_\nu
(k_\rho-1)}{\tilde\nu k_T}\right]^{1/(k_T+k_\rho-1)}\; .
\label{eq:rmhigh}
\end{equation}
Physically, this means that the high-frequency photons come from
a location in the shell that is due to a competition between the
optical depth and the temperature gradient, with the temperature
gradient driving $\tilde{r}_{m}(\nu)$ inwards, while the optical depth
drives $\tilde{r}_{m}(\nu)$ outwards.  

In principle, this expression for the shell radius can give
$\rt_{m\,\rm high}>\rct$, whereas in fact
$\tilde{r}_{m}\rightarrow\rct$ as the shell envelope becomes
very opaque.  Thus, in general the shell radius is given by
\begin{equation}
\tilde{r}_{m,\,\rm high}={\rm Min}(\rct,\tilde{r}_{m,\,\rm high,
	\; steep}) \; . 
\end{equation}

It is important to note that, in finding the contribution function for
the high-frequency regime, we made no assumption about the frequency
dependence of the opacity curve, and thus we may use a tabulated,
realistic opacity curve to find $\tilde{r}_{m}$ for the high-frequency
regime.  We have assumed that $\kappa_{\nu}$ is independent
of position for simplicity, but in principle equation
(\ref{eq:rmhigh}) could be generalized to include
temperature dependent opacity, $\kappa_{\nu}[T(r)]$.

 Evaluating the integral in equation (\ref{eq:steep}) by
the method of steepest descent, we find
\begin{equation}
L_{\nu,\,\rm high}=16 \pi^2(k_{\rho}-1)
\rch^{2}\tilde{\kappa}_{\nu}\tilde{\nu}^{3}\left(\frac{2h\nu_{ch}^{3}}
{c^{2}}\right)\frac{\tilde{r}_{m,\,\rm high}^{2-k_{\rho}}(2\pi/h_{m}'')
^{1/2}
e^{-h_m}
}
{1+2\tilde{\kappa}_{\nu}
\left(\frac{k_{\rho}-1}{k_{\rho}+1}\right)
(\tilde{r}_{m,\,\rm high}^{1-k_{\rho}}-\tilde{r}_{m,\,\rm high}^{2}
\rct^{-k_\rho-1})} \; ,
\label{eq:high_freq_exp}
\end{equation}
where $h_m\equiv h(\rt_{m,\,\rm high})$ and
\begin{equation}
h_{m}''=\tilde{\nu}k_{T}(k_{T}-1)\tilde{r}_{m,\,\rm high}^{k_{T}-2}+k_{\rho}(k_{\rho}-1)\tilde{\kappa}_{\nu}\tilde{r}_{m,\,\rm high}^{-k_{\rho}-1}\; .
\label{eq:hm_double_prime}
\end{equation}
We can express the luminosity in the form of 
the shell model
(eq. \ref{eq:shell})
if we identify the thickness as
\begin{equation}
\Delta\tilde{r}_{m,\,\rm high}=\frac{(2\pi/h_{m}'')^{1/2}}
{1+2\tilde{\kappa}_{\nu}\frac{(k_{\rho}-1)}{k_{\rho}+1}
\left(\tilde{r}_{m,\,\rm high}^{1-k_{\rho}}-
\tilde{r}_{m,\,\rm high}^{2}\rct^{-k_{\rho}-1}\right)}\; .
\end{equation}
The width of the high-frequency gaussian is given by
$(2/h_{m}'')^{1/2}$.  We require $h_{m}''$ $>1$ for proper application
of steepest descent. This condition breaks down at lower frequencies, but we
avoid this problem by combining the high-frequency results
with the low and intermediate ones.  We also note that one can show from 
eqn. (\ref{eq:high_freq_exp}) that our high frequency expression has a 
``super-Wien'' behavior, i.e., it falls off faster than $\nu^{3} \exp(-h\nu/kT)$.

\section{Joint Spectrum}
\label{S:joint}

	To construct a joint spectrum that is valid in all
three frequency regimes, we use the shell model.
In order to recover the limiting cases for the shell
radius and thickness, we sum the results from the low-intermediate
and high frequencies,
\beq
\tilde{r}_{m}={\rm Min}\left\{\frac{\rct C^{1/k_{T}}}
{\rct\tilde{\nu}^{1/k_{T}}+C^{1/\kt}}
+\left[\frac{(k_{\rho}-1)\tilde{\kappa}_{\nu}}
{\tilde{\nu}k_{T}}\right]^{1/(\krho+k_{T}-1)},\; \rct\right\}\; ,
\label{eq:rmp}
\eeq
\beqa
\lefteqn{
\Delta\tilde{r}_{m}=\left[\frac{\Gamma\zeta
\exp\left(\nut \rt_{m,\,\rm low-int}^\kt\right)
}
{(3-k_{\rho}-k_{T})\tilde{\nu}\Gamma\zeta\rct^{-(3-k_{\rho}-k_{T})}+k_{T}
\tilde{\nu}^{(3-k_{\rho})/k_{T}}}\right]\frac{1}
{\tilde{r}_{m,\,\rm low-int}
^{2-k_\rho}}}\nonumber \hspace{3cm}\\
&&+\frac{(2\pi/h_{m}'')^{1/2}}{1+2\tilde{\kappa}_{\nu}
\left(\frac{k_{\rho}-1)}{k_{\rho}+1}\right)(\tilde{r}_{m,\,\rm high}
^{1-k_{\rho}}-
\tilde{r}_{m,\,\rm high}^{2}\rct^{-k_{\rho}-1})}\; ,
\eeqa

	How do these results behave in the limiting cases of
very extended ($\rct\rightarrow\infty$) and very
compact ($\rct\rightarrow 0$) envelopes? For extended
envelopes, the break frequency, 
$\nu_{\rm break}\rightarrow 0$,
so that the low-frequency regime disappears. Correspondingly,
the shell radius at low and intermediate frequencies
approaches the intermediate value, $\rmti\rightarrow (C/\nu)^{1/\kt}$.
The high-frequency emission changes in this limit only
insofar as $\tau_\nu$ increases somewhat.

In the opposite limit of large optical depths,
($\rch\rightarrow\infty$, so that
$\rct\rightarrow 0$), the photosphere approaches the
cloud surface and the slope of the temperature profile becomes 
very steep ($\kt\gg 1$).  
This case is difficult to realize in
practice since we have assumed that there is
no emission beyond $R_c$, and it is difficult, though 
not impossible, to
realize such sharp boundaries in practice. A possible
example of such a sharp boundary is a photoevaporating
globule (Bertoldi \& McKee 1990).  We do not attempt to treat the $\rct\rightarrow 0$ limit in
this paper, but instead confine our attention to $\rct\ga 1$.

Figure 3a depicts the analytic SED (crosses) overplotted on the
numerical SED (solid line) produced from DUSTY, with the three
frequency regimes marked on the plot.  This is a typical SED from the
region of the physical parameter space labeled as ``high mass
protostar'' in Figure 1, with $L/M\sim \;400 L_{\odot}/M_{\odot}$ and
$\Sigma\sim 1~ \rm g\;cm^{-2}$.  Figure 3b shows the combined contribution
function. Figure 3c is a plot of the opacity curve, WD01's $R_{V}=5.5$.
One should read these three plots left to right, i.e., follow the
marked regions in the SED plot in Figure 3a and correlate them with
the marked regions in the contribution function in Figure 3b.  The
spectral features in the contribution function in Figure 3b correlate
with the spectral features in the opacity curve as depicted in Figure
3c.  For example, the 10 $\mu$m ($3\times 10^{13}$ Hz) increase in the
opacity translates to a corresponding increase in $\tilde{r}_{m}$, as
the $\tau=1$ surface at this frequency is driven outwards, while the 5
$\mu$m ($6\times 10^{13}$ Hz) decrease in the opacity 
causes $\tilde{r}_{m}$ to move 
inwards.

\section{Temperature Profile}

Numerical radiative transfer schemes solve for the temperature profile
within the envelope by enforcing the condition of radiative
equilibrium within every resolution element, i.e., that the total
energy absorbed by a differential volume element equal the total
energy emitted.  This condition of energy balance is equivalent to the
condition of zero flux divergence when the radiation field is
time-independent (Mihalas, p.48).  In particular, in spherical
geometry, this condition is equivalent to the constancy of luminosity
as a function of radius.  Approximate numerical radiative transfer
schemes, such as that of Adams \& Shu (1985), have enforced radiative
equilibrium at a discrete number of points, using a particular form of
the temperature profile, and shown that the total luminosity
transported is approximately constant as a function of radius.
Subsequently, more precise numerical radiative transfer solutions have
solved for the temperature variation in the envelope in full
generality, and enforced radiative equilibrium at fine resolution
intervals, thereby guaranteeing virtually exact constancy of
luminosity (IE97).  

Here, in our analytic treatment of the radiative transfer problem, we
have found that the far-IR emission can be inferred with good accuracy from 
a single power-law temperature profile.  We determine the slope of this temperature profile from the self-consistency condition that the input luminosity exactly equal the
emergent 
luminosity:
\beq
L\equiv 4 \pi \rch^{2}\sigma \tch^{4}\tilde{L}\equiv \int  L_\nu\;
d\nu\; .
\end{equation}
Inserting the shell expression for $L_\nu$ (eq. \ref{eq:shell})
into this expression gives
\begin{equation}
\tilde{L}(\rct,\tch,k_{\rho})\equiv \frac{60(k_{\rho}-1)}{\pi^4} \int \tilde{\kappa}_\nu
\tilde{\nu}^3\tilde{r}_m
^{2-k_{\rho}}\Delta \tilde{r}_m
\exp\left[-\frac{h\nu}{kT(\tilde{r}_m)}-\tau_{\nu}(r_m)\right]
d\tilde{\nu}\; .
\label{eq:ltil}
\end{equation}
This is analogous to requiring constancy of luminosity at two discrete
spatial points; numerical schemes achieve greater accuracy by
iteratively imposing zero flux divergence over fine resolution
elements.  Since $\rmt$ and $\Delta\rmt$ depend on both $\kt$ and $C$, 
this
equation gives a zeroth-order, self-consistent determination of the
slope of the temperature profile, $k_{T}$, when the fitting parameters
$\lt$ and $C$ are taken to be equal to unity.  
To secure better agreement with the numerical results from DUSTY, we
have simultaneously solved for 
$\kt$, $\lt$, and $C$ by imposing (\ref{eq:ltil}), and required that the
maximum difference between the analytic and numerical
SEDs between 3 mm and 30 \micron\ be as small as possible.  Two of
these parameters ($\kt$ and $\lt$) characterize the temperature
profile and one ($C$) determines the
break frequency separating low and intermediate frequencies.
We find that the parameters $\lt$ and $C$ are indeed of order unity.

	For a given dust opacity, the shape of
the SED, and therefore the values of the three
fitting parameters, depend on three source parameters,
$\rct$, $\tch$ and $\krho$. The dependence on $\tch$ arises from
the opacity, $\kappa_\nu$.
We now recall that the opacity is scale-free longwards of 30 $\mu$m.
If most of the flux is emitted longwards of 30 $\mu$m
(corresponding to 
$h\nu/k=480 K$), then 
$\tau_\nu\propto\tilde\kappa_\nu=\nut^p$
from equation (\ref{eq:tau}).
In order to have most of the flux emitted longwards
of 30~\micron, we require $\tch\la 300$~K (for $\krho=1.1$, the
condition is more stringent, $\tch\la 250$~K).
For lower values of $\tch$, the power law approximation to the opacity 
is valid and our fitting parameters will be independent of $\tch$.
By comparing with DUSTY, we have found that it is possible
to further simplify the functional dependences to 
\beq
\kt=\kt(\rct,\;\krho),~~~~\lt=\lt(\rct),~~~~{\rm and}~~~~C=C(\krho)\;.
\eeq

With these dependences, we find an excellent level of agreement between our
analytic results and the numerical results from DUSTY over a parameter
space that spans low-mass protostars to ULIRGs.  

	How do we expect $\kt$ and $\lt$ to depend on $\rct$?
We may heuristically understand the functional form of $k_{T}$ 
by considering the equation of radiative equilibrium, 
which requires
that dust grains radiate as much as they absorb:
\begin{equation}
\int\kappa_{\nu}B_{\nu}d\nu=\int\kappa_{\nu}J_{\nu}d\nu\; ,
\label{eq:RE}
\end{equation}
where $B_{\nu}$ is the Planck function and $J_{\nu}$ is the mean intensity of the radiation field.

First consider the outer envelope, assumed to be optically thin.
Insofar as we can make the approximation $\kappa_\nu\propto \nu^\beta$
in the LHS of this equation, it 
scales as $T^{4+\beta}$; the RHS scales as $1/r^{2}$ in the limit of
negligible optical depths.  Therefore, in the outer
parts of extended, optically
thin envelopes, we have
$T(r)\propto r^{-2/(4+\beta)}$. If the envelope is cool
($\tch\la 100$~K), we have $p\simeq 2$ so that
the slope of the temperature profile
$\rightarrow \frac 13$. In fact, extended envelopes,
which have large values of $\rct$, generally have
higher temperatures, so that the effective value of $p$
is somewhat less than 2 and the slope is somewhat greater than $\frac
13$. Next consider the inner, opaque region of the envelope.
There the diffusion approximation is appropriate, and it
gives $T(r)\propto r^{-(1+k_{\rho})/(4-\beta)}$,
which is steeper than the slope in the outer envelope. 
For very opaque envelopes, the gradient can become
steeper than this near the photosphere, just as in
the case of a stellar atmosphere.
We wish to choose a single power
law, $\kt$, to represent the temperature gradient near and outside
the photosphere. 
For extended envelopes (large $\rct$) we expect $\kt$ to
approach $2/(4+\beta)$, whereas for compact envelopes
(small $\rct$) we expect $\kt$ to increase as $\rct$
decreases. We find that this expected variation
can be represented by the sum of two inverse power laws,
\begin{equation}
k_{T}=\frac{A}{\rct^{n_{1}}}+\frac{B}{\rct^{n_{2}}}\; ,
\end{equation}
where $n_{1}$ and $n_{2}$ are positive numbers, 
with $n_{1}$ being a small power in order to represent
the dependence in extended envelopes.  

	The normalization of the temperature profile is
regulated by $\lt$, since
\begin{equation}
\tch\propto\tilde{L}^{-\frac{(k_{\rho}-1)}{2\beta+4(k_{\rho}-1)}}
\label{eq:T_correct}
\end{equation}
from equation (\ref{eq:tch}). This dependence is weak, but
enables us to improve the accuracy of our fits - analogous to a 
temperature correction procedure.
We may understand the variation of $\tilde{L}$ with $\rct$ as
describing the transition from a 
modified
blackbody to a protostellar envelope:
In the limit of large optical depths, 
(small $\rct$)
$\tilde{L}$ is about 1 since 
most of the emission comes from the vicinity of a Rosseland photosphere.
On the other hand,
as $\rct$ tends to infinity, 
the photosphere is less well-defined.
In this limit, $\tilde{L}$ is larger than unity,
reflecting the effective increase in the total emitting area, as the
intermediate frequencies probe the source function over the extended
envelope. We depict the contribution functions for these two
limiting cases in Figure 4.  For large
optical depths, $\tilde{r}_{m}$ is 
approximately constant, since
most of the
emission comes from the surface and the photosphere is relatively
well defined. On the other hand,
at large $\rct$ there is a wide range of $r_m$ 
in the intermediate frequency regime.
 We find that this behavior can be represented by
\beq
\tilde{L}=A_L\rct^a\; ,
\label{eq:ltil1}
\eeq
with sufficient accuracy.

To determine the 
values of $k_{T}$, $\tilde{L}$,
and $C$ for given values of $\rct$ and $\krho$,
we use the downhill simplex method of Nelder and Mead 
(1965) to minimize the largest error between the 
shape of the normalized analytic spectrum and the normalized DUSTY 
spectrum.  We ran DUSTY at an intermediate value 
of $L/M$ ($=40 L_{\odot}/M_{\odot}$), for $k_{\rho}=
1.1,\;1.5$, and 2, increasing $\Sigma$ until we reached $\tch=200 
\rm K$.  To avoid calibrating our functional forms of $k_{T}$,
$\tilde{L}$, and $C$ with respect to higher values of $\tch$, we then 
moved down the $\tch=200 \rm K$ isotherm towards large $\rct$ until we
reached $\Sigma=0.03$. Along this line of 
constant $L/M$ and this isotherm, we have ensured that the input
luminosity equals the emergent luminosity. 
This is not exactly true away from this trajectory due to the
weak dependence of the parameters on $\tch$ that we have
ignored. 

	The temperature profile is determined by $\kt$ and $\lt$.
Our results for the slope of the temperature 
profile are:
\begin{equation}
k_{T}=\frac{0.48k_{\rho}^{0.05}}{\rct^{0.02k_{\rho}^{1.09}}}
+\frac{0.1k_{\rho}^{5.5}}{\rct^{0.7k_{\rho}^{1.9}}}\; .
\end{equation}
Figure 5 depicts $k_{T}$ as a function of $\rct$.
For $\lt$ we find that
\begin{equation}
\tilde{L}=0.87\rct^{0.084}  \; 
\label{eq:ltil_val} 
\end{equation}
is sufficiently accurate for all $\krho$. 

We find that setting $C=\rm constant$ 
for each value of $\krho$ allows for sufficient accuracy (in fact,
letting $C$ vary as a function of $\rct$ does not increase accuracy).   
Values of
$C(k_{\rho})$ are selected to give
good agreement with the break frequency for large $\rct$
envelopes.  This gives $C=1,0.9,$ and $0.5$
for $k_{\rho}=1.1,\; 1.5$ and 2 respectively.  
For other values of $\krho$ in the range $1\leq\krho\leq2$, 
$C$ can be found from
\beq
C=0.27+1.3\krho-0.6\krho^2.
\label{eq:C_eq}
\eeq
A more detailed discussion of this factor
and its determination from the spectrum itself, i.e., the 
break frequency, for extended envelopes is given in \S 8.

We conclude with a caveat on our temperature specification
procedure.  The near-IR flux depends sensitively on the temperature
profile, and we cannot recover the near-IR flux accurately with 
a single power law for the temperature profile.  
For example, as shown in Figure 3, it is clear that
the emission at $5 \mu\rm m$, where
there is an opacity minimum, originates inside the
characteristic radius.  
Our single power law underestimates the
temperature, and therefore the emission,
for regions well inwards of the photosphere.  
This problem could be remedied with a two-component power law for the
temperature profile, but that is not necessary here since we are
focusing on emission at longer than near-IR wavelengths.

\section{The Analytic SED and Its Accuracy}
\label{S:accuracy}

	For convenience, we collect the equations that we have derived
for the far-IR SED of dusty sources:
The mm to far-IR SED is given by:
\begin{equation}
L_{\nu}=16\pi^2(k_{\rho}-1)
\rch^{2}\left(\frac{2h\nu_{ch}^{3}}{c^{2}}\right)
\tilde{\kappa}_{\nu}\tilde{\nu}^{3}\tilde{r}_{m}^{2-k_{\rho}}
\Delta\tilde{r}_{m}\exp\left[-\frac{h\nu}{kT(\tilde{r}_{m})}-
\tau_\nu(\tilde{r}_{m})\right]\; .
\end{equation}
The characteristic emission radius, $\tilde{r}_{m}=\tilde{r}_{m}(\nu)$, is the location in
the shell where most of the flux in a given frequency-band originates from, (the ``m'' is for maximum), and is given by:
\beq
\tilde{r}_{m}={\rm Min}\left
(\tilde{r}_{\rm m,low-int}+\tilde{r}_{\rm m,high}, 1 \right)\; ,
\eeq
where the total $r_{m}$ is the sum of the high frequency $r_{m}$ and
the combined low-intermediate frequency $r_{m}$,
\beq
\tilde{r}_{\rm m,high}=\left[\frac{\tilde\kappa_\nu
(k_\rho-1)}{\tilde\nu k_T}\right]^{1/(k_T+k_\rho-1)}\; ,
\eeq
\beq
\tilde{r}_{\rm m,low-int}=\frac{\rct
C^{1/\kt}}{\rct\tilde{\nu}^{1/k_{T}}+C^{1/\kt}}\; .
\eeq
The shell thickness, $\Delta\tilde{r}_{m}$ is given by:
\beqa
\lefteqn{
\Delta\tilde{r}_{m}=\left[\frac{\Gamma\zeta
\exp\left(\nut \rt_{m,\,\rm low-int}^\kt\right)
}
{(3-k_{\rho}-k_{T})\tilde{\nu}\Gamma\zeta\rct^{-(3-k_{\rho}-k_{T})}+k_{T}
\tilde{\nu}^{(3-k_{\rho})/k_{T}}}\right]\frac{1}
{\tilde{r}_{m,\,\rm low-int}
^{2-k_\rho}}}\nonumber \hspace{3cm}\\
&&+\frac{(2\pi/h_{m}'')^{1/2}}{1+2\tilde{\kappa}_{\nu}
\left(\frac{k_{\rho}-1)}{k_{\rho}+1}\right)(\tilde{r}_{m,\,\rm high}
^{1-k_{\rho}}-
\tilde{r}_{m,\,\rm high}^{2}\rct^{-k_{\rho}-1})}\; ,
\eeqa
where the argument $(3-k_{\rho})/k_{T}$ for the Gamma and Zeta 
functions has been suppressed for clarity, and 
$h_{m}''$ is given by Eqn. \ref{eq:hm_double_prime}.   
The ratio of $h\nu$ to $kT$ equals $C$ for the intermediate frequencies and is given by: 
\beq
C=0.26+1.3k_{\rho}-0.6k_{\rho}^{2}
\eeq
and
\beq
k_{T}=\frac{0.48k_{\rho}^{0.05}}{\rct^{0.02k_{\rho}^{1.09}}}
+\frac{0.1k_{\rho}^{5.5}}{\rct^{0.7k_{\rho}^{1.9}}}\; .
\eeq
The SED variables and the source parameters are related by:
\begin{equation}
\rct=\left\{\frac{(L/M)\Sigma^{\frac{4+\beta}{\beta}}}{4\sigma \times 0.87}
\left[\frac{(3-k_{\rho})\kappa_{\nu_{0}}}{4(k_{\rho}-1)T_0^\beta}\right]^{4/\beta}
\right\}^{-\frac{\beta}{1.92\beta+4(k_{\rho}-1)}}\; ,
\end{equation}
\begin{equation}
\tch=\left\{\frac{L/M}{4\sigma \times 0.87
\Sigma^{\frac{2.92-k_{\rho}}{k_{\rho}-1}}}
\left[\frac{4(k_{\rho}-1)T_{0}^{\beta}}{(3-k_{\rho})\kappa_{\nu_{0}}}\right]
^{\frac{1.92}{k_\rho-1}}\right\}^{\frac{k_{\rho}-1}{4(k_{\rho}-1)+1.92\beta}}
\; ,
\end{equation} 
where we have substituted for the values of $A_{L}$ and $a$ from (\ref{eq:ltil_val}).

   The physical parameter space 
in which we have determined the accuracy of our analytic SED, 
as compared
with the numerical solutions from DUSTY, 
is given in Figures 1, 6, and 7
for three density profiles, 
$k_{\rho}=1.1,\;1.5$ and 2.  
The parameter space extends over luminosity-to-mass 
ratios $L/M=(0.1-4000)L_\odot/M_\odot$ and
surface densities $\Sigma=0.01-100$ g cm$^{-2}$,
which encompass the full range of astronomical dusty sources. 
We restrict our attention to $\tch<300$~K (250 K for $\krho=1.1$) so
that most of the emission is longwards of 30 \micron\
and the slope of the temperature profile is
approximately independent of $\tch$. We also restrict
our attention to $\rct>1$, since our assumption of a spherical cloud
with a sharp boundary is likely to break down for smaller envelopes.
Within these boundaries, the SED given by the above equations is
accurate to within a factor 2 at worst.  The largest errors occur for
$\krho=1.1$; for $\krho=1.5$, the SED is accurate to within a factor of
1.6, and for $\krho=2$ it is accurate to within a factor 1.5.
If we focus on surface densities in the range $0.01-3$~g~cm$^{-2}$, which
is indeed where the majority of astrophysical source lie, the
accuracy is 1.5 for $\krho=1.5$ and 1.3 for $\krho=2$.  
We also note that our equations hold
for $k_{\rho}=1$, as numerical runs verify that $k_{\rho}=1$ envelopes
are well approximated by $k_{\rho}=1.1$ (to better than 10 \%) over our parameter space.  
Finally, we note that our
largest errors, 
as compared to the numerical results,
are at 30 $\mu\rm m$, on the low $\rct$ end (large $\Sigma$ region), 
where there is very little flux.  As a result, our accuracy in 
inferring source parameters 
(given that our assumptions stated in \S 2 
are satisfied) 
is generally better than a factor of 1.5, 
a point that discuss in more detail in Paper II.  SED fits, to
a low-mass protostar, massive protostar, and ULIRG, respectively, are given
in Figure 2a, Figure 2b, and Figure 2c.

We note here briefly (see Paper II for a more detailed explanation 
of the application of our methodology to observed data) that 
our accuracy criteria hold only so long as our assumptions are valid.
In particular, 
many of the sources we consider in Paper II have significant high 
frequency fluxes that our solution cannot account for.  
High frequency emission can escape 
due to inhomogeneities in the envelope, due to the presence of 
an accretion disk, or from distributed sources of luminosity in
the field of view that are not obscured by dust.  We consider
the effects of an inhomogeneous dust distribution in Paper III, 
and the effects of distributed sources of luminosity in Paper IV.  
To avoid fitting to data that may predominantly arise from one of these 
additional agents, in Paper II 
we have performed our fits from mm - 60 $\mu\rm m$,
where these additional complexities are unlikely to significantly 
change the emitted spectrum.  If the fit based on the 
mm - 60 $\mu\rm m$ data also fits the high frequency data, that is
a good indication that the high frequency data are not due to these
additional agents.

Our basic assumption of spherical symmetry can be violated by the
presence of a disk that is large and massive enough to influence 
the far-IR spectrum.   Since the disk radius approximately demarcates the region where the density, and therefore the temperature, profiles
become nonspherical, our
method is applicable only to the case in which the disk radius is smaller than
the characteristic radius, $R_d < \rch$.  Even if this geometric
condition is satisfied, we also require that emission at long wavelengths 
(where the emissivity scales as the product
of the mass times the temperature) from the
disk be small compared to that from the envelope, which implies
\beq
M_{\rm disk} \ll M_{\rm env} \left(\frac{T_{\rm env}}{T_{\rm disk}} 
\right)\; .
\eeq

	We have also assumed that emission from the vicinity of
the dust destruction radius at $\rdd$
does not significantly influence the
far-IR spectrum.  A sufficient condition to ensure this
is that the optical depth to the dust
destruction front at photospheric frequencies be large; we
require $\tau>3$ at the frequency $\nu_{\rm peak}$
at which $\nu L_{\nu}$ peaks, which implies
\beq
\frac{\rdd}{R_{c}} <
\frac{1}{\left(1+3\nut_{\rm peak}^{-p}
\rct^{k_{\rho}-1}\right)^{\frac{1}
{k_{\rho}-1}}} \; .
\eeq

Figure 8 depicts $\nut_{\rm peak}=\nu_{\rm peak}/\nu_{\rm ch}$ as a function
of $\rct$, and \S 8 gives corresponding fits to the curves.  
This optical depth condition is satisfied for $k_{\rho}=2$ and $3/2$ for
all $\rct$ in our parameter space.  
It is also satisfied for $k_{\rho}=1.1$ when $\rct \la
10$.  However, for $k_{\rho}=1.1$ and $\rct\ga 10$, the optical depth
to the dust destruction front at $\nu_{\rm peak}$
is not large.  Nevertheless, emission
from $\rdd$ does not influence the spectrum because the
far-IR luminosity originating from the vicinity of
the dust destruction radius is small compared to that
from the vicinity of the
characteristic radius in our parameter space.   
An upper limit on emission
from the vicinity of the dust destruction front is 
a blackbody radiating at $T_{\rm dd}$.
Hence, at intermediate frequencies, a sufficient condition for
ignoring emission from the dust destruction front is
\beq
T_{\rm dd}R_{\rm dd}^{2} \ll T_{\rm ch}R_{\rm ch}^{2}\exp{(-C)}\; , 
\eeq
(with $h\nu/kT=C$), which is always satisfied in
our parameter space. At frequencies well above $\nu_{\rm peak}$, where
the emission from the vicinity of the dust destruction front is
significant, the optical depth is large enough to suppress it. 

\section{Shape of SED:  $\rct$ and $\tch$}
\label{S:shape}

As long as the assumptions stated in \S 2 \& \S 7 are satisfied,
i.e., given a spherically symmetric, homogeneous distribution of dust
illuminated by a central source of radiation, such that
the emitted spectrum is not significantly affected
by emission from the dust destruction front,
the shape of the long-wavelength
SED will depend only on $\rct$ for a given density profile and dust
model.  Thus, moving along a line of constant $\rct$ preserves the shape of
the SED, while shifting the peak of the SED as one intersects  lines
of different $\tch$.  The ratio of the peak frequency to the
characteristic frequency is shown
in Figure 8 and can be approximated as 
\beq
\frac{\nu_{\rm peak}}{\nu_{\rm ch}} \simeq
0.82k_{\rho}+\frac{5.4-1.8k_{\rho}}{
\rct^{0.56k_{\rho}-0.22}}\; .
\label{eq:nupk3}
\eeq
Similarly, moving along an isotherm, i.e., a
line of constant $\tch$, 
changes the shape of the SED as $\rct$ changes but
preserves near constancy of the peak of the
SED since the variation of $\nu_{\rm peak}$
with $\rct$ is weak over most of the parameter
space (see eq. \ref{eq:nupk3}). 

	Note that the ratio of the peak frequency to the characteristic
frequency decreases 
as $\rct$ increases because the intermediate frequencies, 
which are emitted at lower
temperatures, become relatively more important in the total 
energy balance.  In the low $\rct$ regime (large optical depths), our
results are 
qualitatively similar to the blackbody limit, as $\nu_{\rm peak}\sim 
4\nu_{\rm ch}$ (recall that $\nu_{\rm peak}$ is the frequency at which
$\nu L_\nu$ peaks). 
Over much of the astrophysical parameter space, however, spectra peak 
at $\nu_{\rm peak}\sim (1-1.6)
\nu_{\rm ch}$.

Low mass protostars, as shown in Figure 9a, are
characterized by large $\rct$ and have broad SEDs.  They have 
extended envelopes with temperatures low enough such that the
intermediate frequency region, with its flatter slope, shows a clean
separation from the low-frequency region.  On the other
hand, the far-IR spectra of massive protostars, ULIRGs and super-star
clusters resemble quasi-blackbody spectra, where the low,
intermediate, and high-frequency components have become 
smeared together (Figure 9b).  
Fits to observed data (Figure 2a, 
Figure 2b, and 2c) also display this variation of shape, as $\rct$
varies over the parameter space, from the low-mass protostars 
to the massive protostars. 

We also give the general relation between the characteristic parameters
and the photospheric parameters.  As noted earlier, our definition of
the characteristic parameters coincides with photospheric parameters
when core radius is much larger than $\rch$. However, as $\rct$ 
becomes small,
the Rosseland photosphere moves to the surface of the core.
Therefore, we approximate
the relation between the photospheric and
chracteristic parameters as a harmonic 
mean of $R_{c}$ and $\rch$:
\begin{equation}
R_{\rm ph}=\frac{R_{\rm ch}}{1+\rct^{-1}}\; .
\end{equation}
We use $L=4\pi R_{\rm ph}^{2}\sigma T_{\rm ph}^{4}=4\pi R_{\rm
ch}^{2}\sigma T_{\rm ch}^{4}\tilde{L}$ to see that the photospheric
temperature 
varies as:
\begin{equation}
T_{\rm ph}=\tch\lt^{1/4}(1+\rct^{-1})^{1/2}\; .
\end{equation} 
  
\subsection{The Three-Component Spectrum of Extended Envelopes}

We now consider the limiting case of highly extended envelopes, i.e.,
$\rct\rightarrow\infty$.  We reach this region of astrophysical
parameter space by moving downward and to the left
along an isotherm towards low
$L/M$ and low $\Sigma$, i.e., past low mass protostars towards very
low luminosity cores, or nearly prestellar cores.  Figure 11 depicts a
large $\rct$ envelope and its corresponding contribution function,
with the break frequency marked in both diagrams.  We note that in
this limiting case, since $k_{T}\rightarrow \rm constant$, 
we may infer the
source parameters from $\nu_{\rm peak}$ and $\nu_{\rm break}$ only.
The peak frequency is given
by equation \ref{eq:nupk3}; the break frequency is given in terms of
$L/M$ and $\Sigma$ by
\begin{equation}
\tilde{\nu}_{\rm break}=C
\left\{\frac{(L/M)\Sigma^{(4+\beta)/\beta}}{4\sigma 0.87} 
\left[\frac{(3-k_{\rho})\kappa_{\nu_{0}}}{4(k_{\rho}-1)T_{0}^{\beta}}
\right]^{4/\beta}\right\}
^{\beta k_T/[1.92\beta
+4(k_{\rho}-1)]}\; .
\end{equation}
where $C$ is given in equation \ref{eq:C_eq}.
Figure 10 depicts the factor $C(\rct)$ for 
the three standard density profiles; as shown,
$C\rightarrow \rm constant$ for large $\rct$.  

The clean separation of the intermediate frequency regime in
this limiting case allows us to specify the temperature dependence of
the spectrum
in the intermediate frequency regime,
\begin{equation}
L_{\nu}\propto \tch^{7-3k_{\rho}}\; ,
\end{equation}
which follows from Eqn. \ref{eq:lnu} and Eqn. \ref{eq:int} with $k_{T}=1/3$ and $\beta=2$.  The slope of $k_{\rho}=1.1,1.5$ and 2
envelopes depends on $\tch^{4}$, $\tch^{5/2}$, and $\tch$ respectively, so
that $k_{\rho}=1.1$ envelopes are most sensitive to the temperature.

As we discuss in \S 8.3, while it is not generally
feasible to discriminate 
density profiles from the far-IR spectra of unresolved sources, 
it is possible to do so for extended envelopes.
In this case, we may write down a simple expression for 
the intermediate frequency slope, which holds for 
$1\ga\tilde{\nu}\ga\tilde{\nu}_{\rm break}$:
\begin{equation}
L_{\nu}\propto \nu^{3+\beta-(3-k_{\rho})/k_{T}(\rct)}\; .
\end{equation}
For envelopes that are sufficiently extended ($\rct \ga 2000$), 
one can infer $\krho$ directly from the measured slope in
the intermediate frequency regime.
The pristine environments of isolated, low-luminosity 
cores allow for the most robust determination of the density profile 
from the far-IR SED, without requiring resolved observations of 
the source itself.  Such a determination of the density profiles 
for these sources would also provide a direct observational 
test of the star formation scenario, as different 
theories of star formation are characterized by different density profiles.

\subsection{Density Profile}

The inference of density profiles from the far-IR SED has been treated in
various ways in the literature.  Some authors have determined the
density profile from fitting to the SED and the intensity profile
concomitantly (Shirley et al 2002, Mueller et al 2002, Beuther et al
2002), which works if the
intensity is solved for self-consistently.  The Adams (1991)
approximation for the spatial distribution of the millimeter and
submillimeter emission, and its use to infer density profiles
(e.g. Beuther et al 2002), is based on the assumption that $1/3\la
k_{T}\la 2/5$, which is a good approximation for large $\rct$.
However, the dense envelopes of massive protostars imply they are
characterized by lower values of $\rct$, and therefore higher values
of $k_{T}$.  Hence, the accuracy of the Adams approximation for dense
envelopes is degraded in this region of the parameter space.  Some
authors (e.g. van der Tak et al 2000, OLD99) have claimed that density
profiles can be inferred from the observed SED alone.  Others,
(e.g. Correia et al 2004) have noted that varying density profiles
give equally good fits to the SED.  Therefore, the question remains:
under what conditions can the density profile be inferred from the mm
to far-IR SED alone?

So far, we have shown that for a given density profile and dust model,
the SED is a function only of $L/M$ and $\Sigma$.  To address the question
of whether inference of density profiles is feasible, we now
ask under what conditions can one distinguish
the predicted mm to far-IR fluxes for envelopes
characterized by different density profiles?
To do this, we consider SEDs produced by envelopes characterized by
the same $L/M$ and $\Sigma$, but with different density
profiles. Figure 12a depicts the the $\rm 1 mm$ to $\rm 60
\mu m$ flux ratio vs. the the $\rm 30 \mu m$ to $\rm 60 \mu m$ flux ratio
for the three density profiles, $k_{\rho}=1.1,3/2,2$, with each trio
having the same $L/M$ and $\Sigma$ values.  The $k_{\rho}=3/2$ points
(crosses) correspond to an isotherm, $T_{\rm ch}=210$~K, 
with the
bottommost point corresponding to $\rct=43$ and the topmost point to
$\rct=370$.  Note that for the first two trios, the colors for
the different density profiles are not clearly distinguishable 
(i.e., they do not differ by more than a 
a factor of two), while for the third trio ($\rct=180,\tch=210$~K for
$k_{\rho}=3/2$) they are.  Thus, 
the different density profiles are distinguishable at
sufficiently large $\rct$. The region in the $L/M-\Sigma$ plane
in which at least one of the two colors differs by at least
a factor 2 between the $\krho=1.1$ and $\krho=2$ cases is
shown in Figure 12b.

The basic intuition behind this result is
that envelopes with large values of $\rct$ 
(i.e., envelopes in which the photosphere is small compared to
the size of the core)
have an extended
range of intermediate frequency emission in the far-IR.
Emission at intermediate frequencies originates from
a range of radii and temperatures, and as a result, 
the emission from large $\rct$ envelopes depends more
sensitively on the density profile than that from
low or moderate $\rct$ envelopes.
This is true for any value of $\tch$, i.e., 
it will be easier to discriminate density profiles for larger values
of $\rct$ whether we consider sources at high $\tch$ or 
low $\tch$.  However, as shown
in Figure 12b, there is some dependence on $T_{\rm ch}$, 
since the ratio of the 30 \micron\ emission to 60 \micron\ emission
is sensitive to the temperature; this effect is ameliorated
by the fact $\nu_{\rm peak}/\nu_{\rm ch}$ increases at 
low values of $\rct$, where $\tch$ is low also.
  
We conclude that while it may be feasible
to infer the density profile from the SED for low and
intermediate mass protostellar sources, it is 
difficult to do so for high-mass protostars and extragalactic sources
as they are generally on the low $\rct$ end of the parameter space.
  
\section{Conclusion}

We have presented an analytic, self-consistent solution for
the spectral energy distribution of homogenous, spherically symmetric,
dust-enshrouded central sources of radiation that are not affected by
emission from the dust destruction
radius.  The main points are:

1. For a given dust model and density profile, the SED is determined
by three parameters, the luminosity of the central source,
$L$, the mass of the envelope, $M$, and the size of the envelope,
$R_c$. The {\it shape} of the SED is distance independent, and
is determined by the two distance-independent parameters
$L/M$ and $\Sigma\equiv M/(\pi R^2)$. The cases of greatest
relevance in astrophysics have $0.1L_\odot/M_\odot 
\la L/M\la 4000 L_\odot/M_\odot$ and $0.01\; {\rm g~cm^{-3}}\la\Sigma
\la 100\;{\rm g~cm^{-3}}$. We consider power-law density
profiles, $\rho\propto r^{-\krho}$, with $1\leq\krho\leq 2$.

2. The characteristic radius, $\rch$, is analogous to a Rosseland
photosphere.   The characteristic temperature, $\tch$, is given
by $L=4\pi\lt \rch^2\sigma \tch^4$, where $\lt$ is a number
of order unity. 
The emission at a frequency $\nu$ can be viewed as originating from
a shell centered at $r_m(\nu)$, with thickness $\Delta r_m(\nu)$
and with an attenuation 
$e^{-\tau(\nu)}$.  Most of the emission
originates outside $\rch$, since for $h\nu\ga k\tch$ the
optical depth is large, whereas 
for $h\nu\la k\tch$ most of the emitting mass is outside
the photosphere.  

3.  The SED is readily described in terms of the characteristic
temperature, $\tch$, and the dimensionless parameter
$\rct\equiv R_c/\rch$. For a given dust model and density
profile, $\tch$ and $\rct$ are unique functions of $L/M$ and
$\Sigma$. Low-mass protostars have extended
envelopes with core radii that are large compared to
the photospheric radius, so that $\rct\gg 1$. High-mass protostars,
ULIRGs and super--star clusters generally have less extended
envelopes (smaller values of $\rct$). The width of the SED correlates
with $\rct$: sources with small values of $\rct$ have quasi-blackbody
SEDs, whereas those with large values of $\rct$ have broad SEDs.

4. The SED has three frequency regimes. 
Low frequencies are emitted by the entire envelope, and
have $h\nu< kT(R_c)$. For the density profiles we have considered,
most of the low-frequency emission comes from the outer
part of the envelope; it is optically thin. 
Intermediate frequencies have $kT(R_c)<h\nu\la k\tch$, so that
they are suppressed in the outer envelope; most of the emission is
optically thin. Finally, high frequencies have $h\nu\ga k\tch$.
The flux at high frequencies is the result of a competition
between the opacity, which favors emission from larger
radii, and the temperature, which favors emission from smaller
radii, where the dust is warmer.  Low frequencies vary as
$\nu^{2+\beta}$, intermediate frequencies vary as $\nu^{3+\beta-(3-k_{\rho})/k_{T}(\rct)}$, and high frequencies have a ``super-Wien'' behavior - they 
fall off faster than $\nu^{3}\exp(-h\nu/kT)$.

5.  We approximated the temperature profile near the photosphere as a
power law, $T\propto r^{-k_{T}}$, where the temperature gradient,
$k_{T}$, is determined from imposing the self-consistency condition
that the input luminosity exactly equal the emergent luminosity.  We
then showed that adoption of this approximate temperature profile
leads to good agreement 
(usually within a factor $\sim 1.5$ between 3
mm and 30 $\mu$m) with spectra calculated with the numerical
radiative transfer code DUSTY.

6. Longwards of 30 $\mu$m, where the opacity is scale free, the shape of
the SED is a function only of $\rct$ and $\krho$.  
However, the dependence on the density profile ($\krho$) is significant
only for relatively large values of $\rct$.
Large $\rct$ envelopes, such as those of low-luminosity, low-mass protostars,
 present an opportunity to infer the source parameters from two observable frequencies, $\nu_{\rm peak}$ and $\nu_{\rm break}$, 
where $\nu_{\rm break}$ divides the low and intermediate
frequency regimes. At large $\rct$, the intermediate 
frequency regime has a slope that depends explicitly on 
$\krho$.

We thank Bruce Draine, Eugene Chiang, David Hollenbach, Xander
Tielens, and especially Gibor Basri for helpful discussions.  We also
thank Maia Nenkova and Dejan Vinkovic for answering innumerable
DUSTY-related questions.  This research is supported in part by NSF
grant AST00-98365.  CFM gratefully acknowledges the support of the
Miller Foundation for Basic Research.  SC gratefully acknowledges a
dissertation year fellowship from the 
Graduate Opportunity Program.

\appendix
\section{Relation to DUSTY Parameters}
We give the relations between the characteristic parameters and the
DUSTY parameters.  First, we note that the DUSTY dimensionless
parameters are: 
$y=R_{c}/\rdd=\rct/\rddt$; $\tau_{\nu_{0}}$, 
the optical depth through the shell at frequency $\nu_0$; 
and $k_{\rho}$, 
the slope of the density profile.  
The dimensional parameters are the dust destruction temperature, 
$T_{\rm dd}$, and the temperature of the central illuminating source, 
$T_*$.  In addition, the shape of the opacity curve is 
required as input.  The dust destruction radius, $\rdd$, can be expressed as
\begin{equation}
\rdd=\left[(1+0.007\tau_{V})\frac{\bar Q_{\rm UV}}{\bar Q_{\rm IR}}\;
\frac{L}{16\pi\sigma T_{\rm dd}^{4}}
\right]^{1/2}\; ,
\end{equation}
as in IE97.

Thus, the value of the dust destruction radius depends explicitly 
on the magnitude of the luminosity.  For our purposes, the 
parameters $T_{\rm dd}$ and $\rdd$ do not enter into the problem, 
as we consider optical depths to the dust destruction radius 
large enough such that it is opaque at far-IR wavelengths.  
We choose $\nu_0$ to be in the power-law portion of the opacity
curve so that we
may express $\tau_{\nu_{0}}$ as
\begin{equation}
\tau_{\nu_{0}}=\left(\frac{h\nu_{0}}{k\tch}\right)^{\beta}\left(
\rddt^{-k_{\rho}+1}-\rct^{-k_{\rho}+1}\right),
\end{equation}
where $\rddt\equiv \rdd/\rch$.
 
\section{Escape Fraction}

Here, we give the derivation of the escape fraction for spherical geometry:
\begin{equation}
f(r)=\int_{\mu=0}^{\mu=1}e^{-\tau_{\nu}(r,\mu)}d\mu\; ,
\end{equation}
where $\mu=\cos(\theta)$.  For escape at small angles (which is appropriate for large optical depth), we evaluate $\tau(\theta)$ for $\theta \ll 1$.  We express the optical depth as:
\begin{equation}
\tau_{\nu}=\tilde{\kappa}_{\nu}(k_{\rho}-1)\int r'^{-k_{\rho}}d\tilde{s}\; ,
\end{equation}
where the path length is equal to
\begin{equation}
ds=\frac{r'dr'}{[r^{2}+r'^{2}(\mu^{2}-1)]^{1/2}}\; ,
\end{equation}
with $r$ as the source point.
We evaluate the optical depth for $\theta \ll 1$, to find that:
\begin{equation}
\tau_{\nu}(r,\theta)=\tau_{\nu}(r,0)+\frac{\frac{1}{2}\theta^{2}r^{2}\tilde{\kappa}_{\nu}(k_{\rho}-1)}{k_{\rho}+1}(\tilde{r}^{1-k_{\rho}}-\tilde{r}^{2}\rct^{-1-k_{\rho}})\; .
\end{equation}
Now, we evaluate the integral for the escape fraction with respect to angle and find that:
\begin{equation}
f(r)=\frac{(k_{\rho}+1)e^{-\tau_{\nu}(r,0)}}{2\knut(k_{\rho}-1)(\tilde{r}^{1-k_{\rho}}-\tilde{r}^{2}\rct^{-1-k_{\rho}})}, ~~~~~~\tau_{\nu}(r,0) \gg 1 \; .
\end{equation}
We assume that the $\tau_{\nu}(r,0)=1$ surface is
at $r \ll R_{\rm c}$, so that the escape fraction is unity near the surface
of the core.  We recover the proper limit when $r=R_{c}$, i.e., 
we recover $f=1$ in that case, if we modify this to be: 
\begin{equation}
f(r)=\frac{e^{-\tau_{\nu}(r,0)}}{1+2\tilde{\kappa}_{\nu}
\left(\frac{k_{\rho}-1}{k_{\rho}+1}\right)\left(\tilde{r}^{1-k_{\rho}}-
\tilde{r}^{2}\rct^{-k_{\rho}-1}\right)}\; .
\end{equation}

\vfill\eject
\begin{figure}[h] \begin{center}
\centerline{\psfig{file=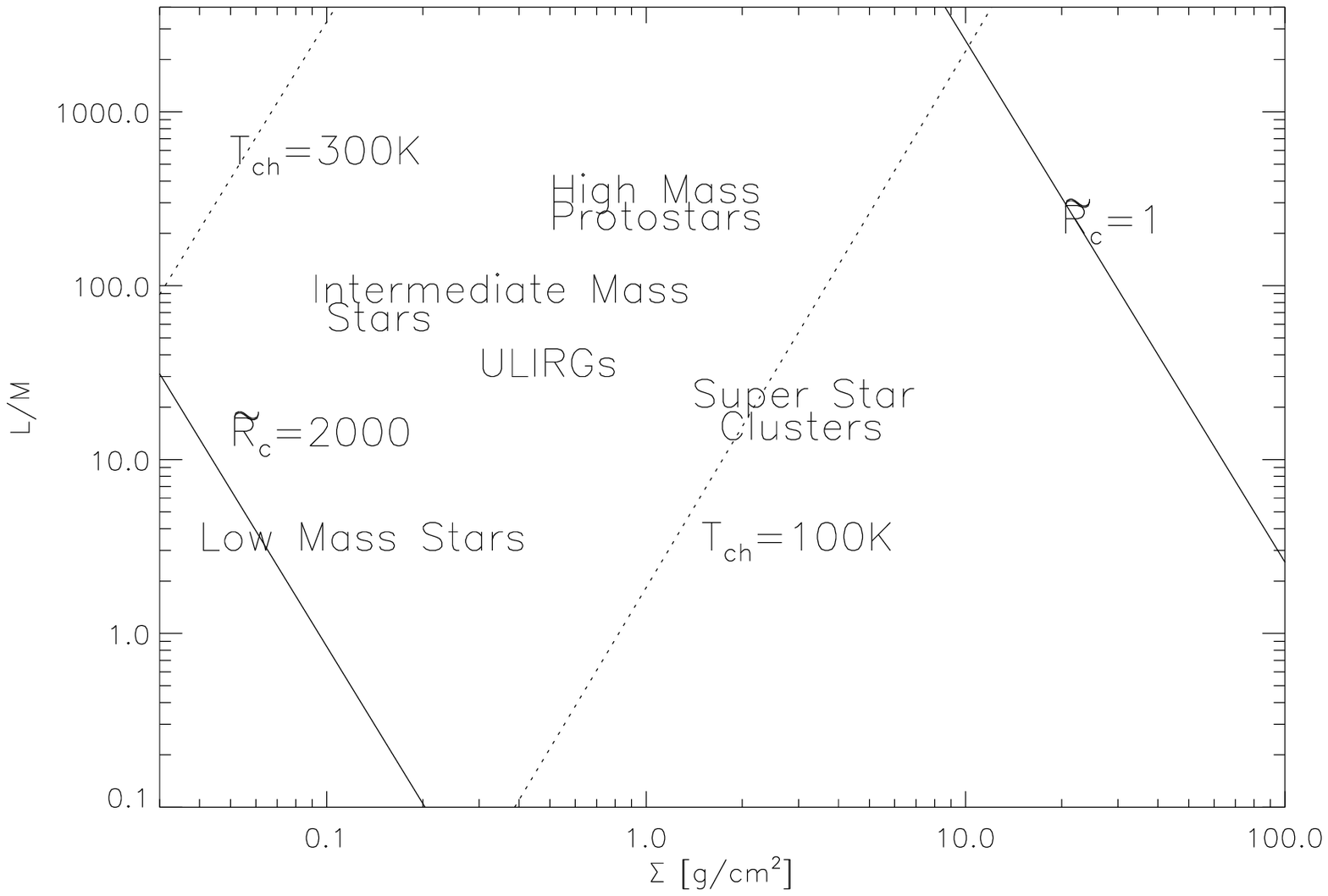,height=3.5in,width=3.5in}
\psfig{file=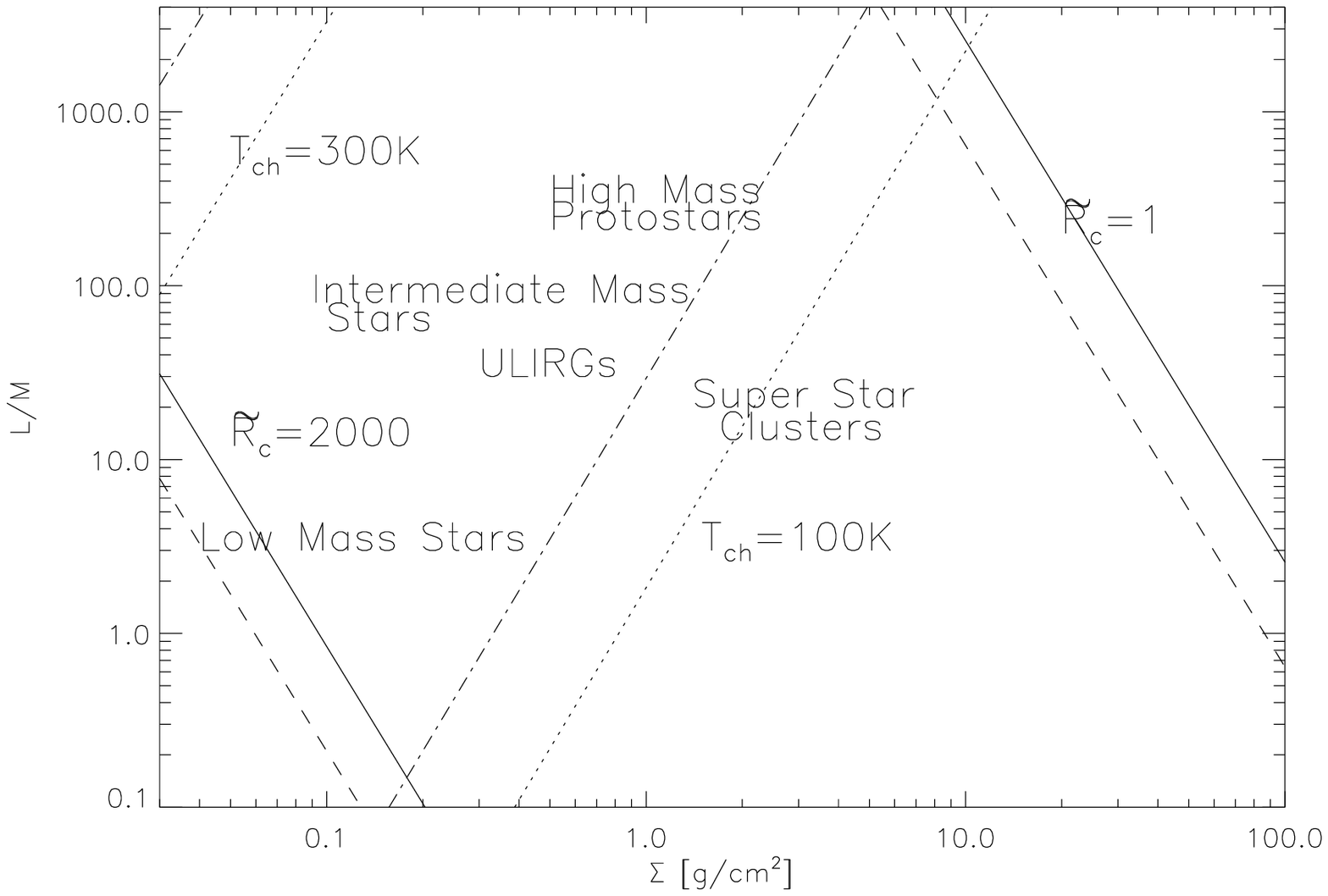,height=3.5in,width=3.5in}}
\end{center}
\caption{(a)$L/M [L_{\odot}/M_{\odot}]$ vs $\Sigma [\rm g/cm^{2}]$ for fiducial density profile, $k_{\rho}=3/2$.  $\tch$ is the effective photospheric
temperature and $\rct$ is the ratio of the core radius to $R_{\rm ch}$, which
is like the Rosseland photosphere  (b) with change in opacity normalization of a factor of two (dashed line), due to ice-mantles}
\end{figure}

\begin{figure}[h] \begin{center}
\centerline{\psfig{file=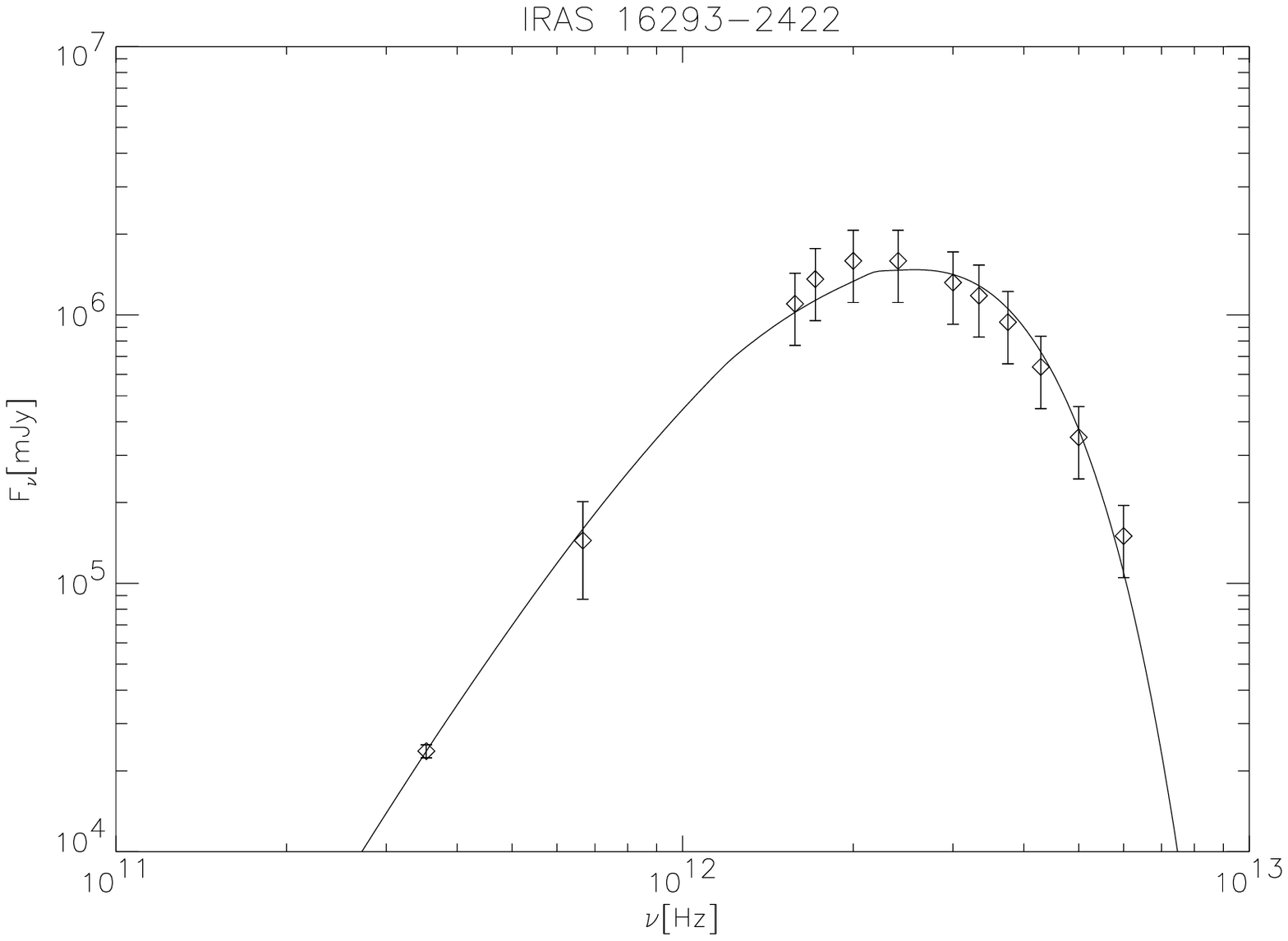,height=2.8in,width=2.8in}
\psfig{file=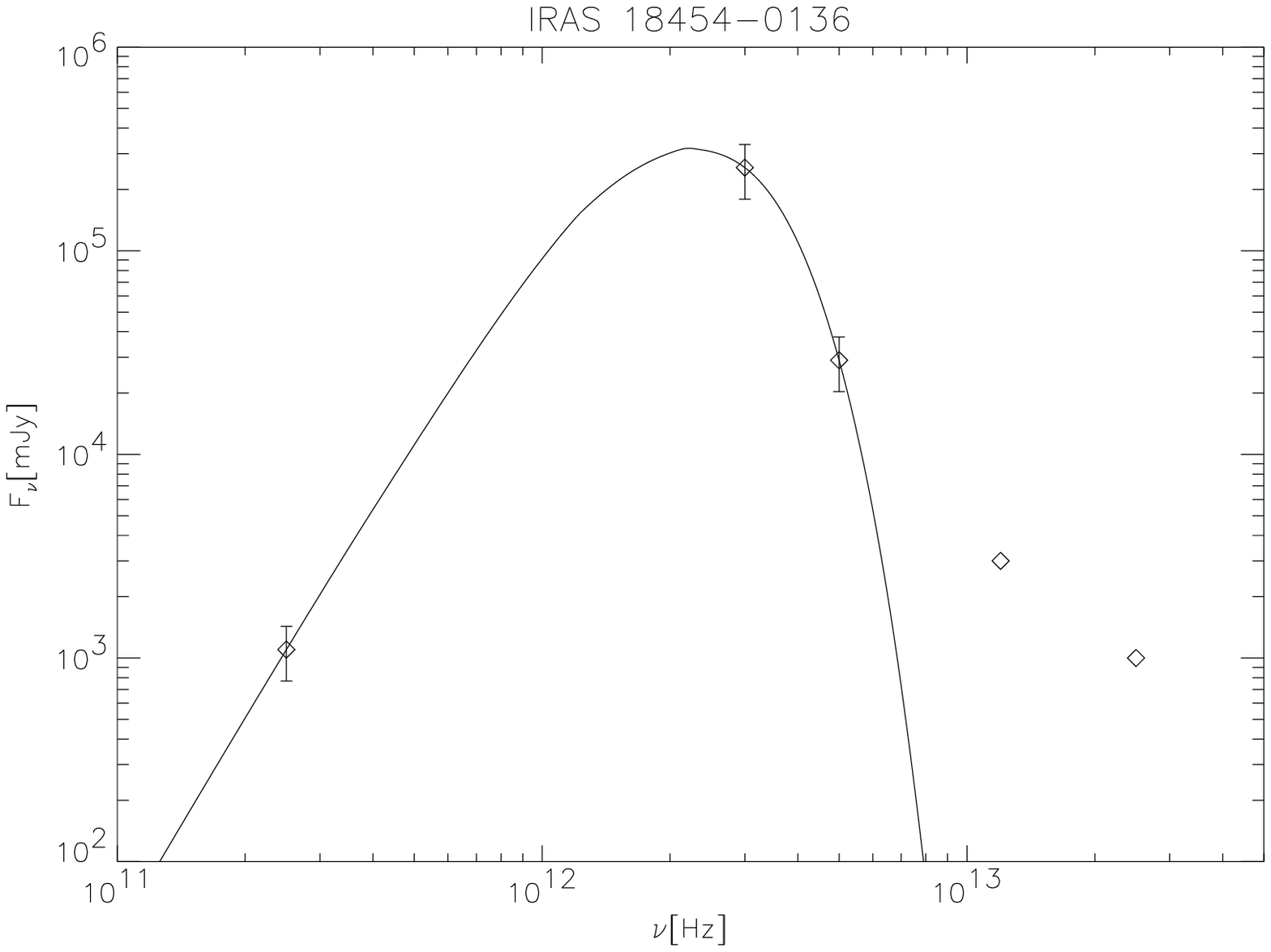,height=2.8in,width=2.8in}
{\psfig{file=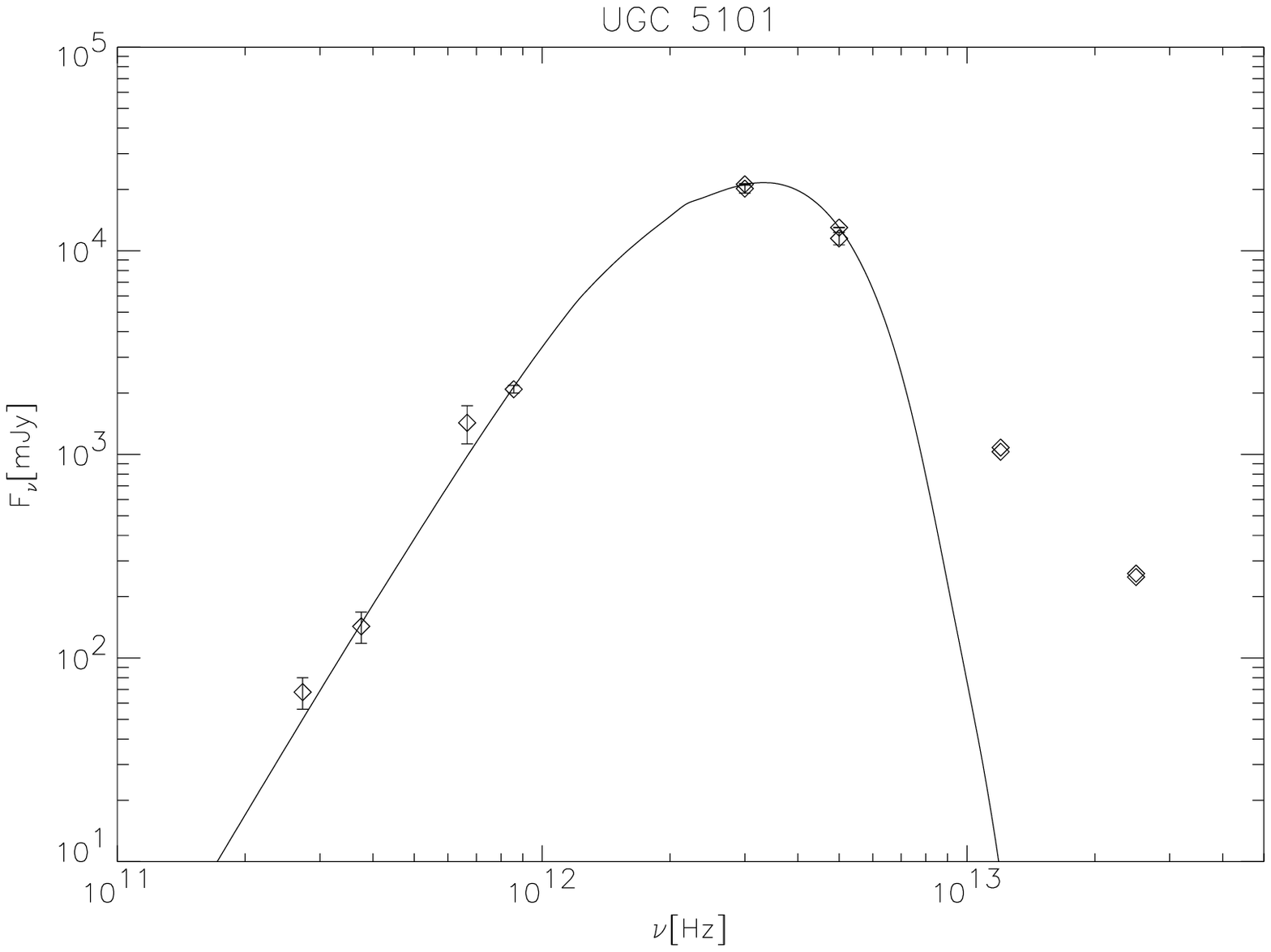,height=2.8in,width=2.8in}}}
\end{center}
\caption{(a) SED fit to low-mass protostar, $L/M \sim 7 L_{\odot}/M_{\odot}$, $\Sigma \sim 0.3 \rm g~cm^{-2}$ (b) SED fit to massive protostar, $L/M \sim 11 L_{\odot}/M_{\odot}$, $\Sigma \sim 1.5 \rm g~cm^{-2}$, (c) SED
fit to ULIRG, $L/M \sim 33 L_{\odot}/M_{\odot}$, $\Sigma \sim 0.3 \rm g~cm^{-2}$.  Details of fitting procedure and inferred parameters given in Paper II}
\end{figure}

\begin{figure}[h] \begin{center}
\centerline{\psfig{file=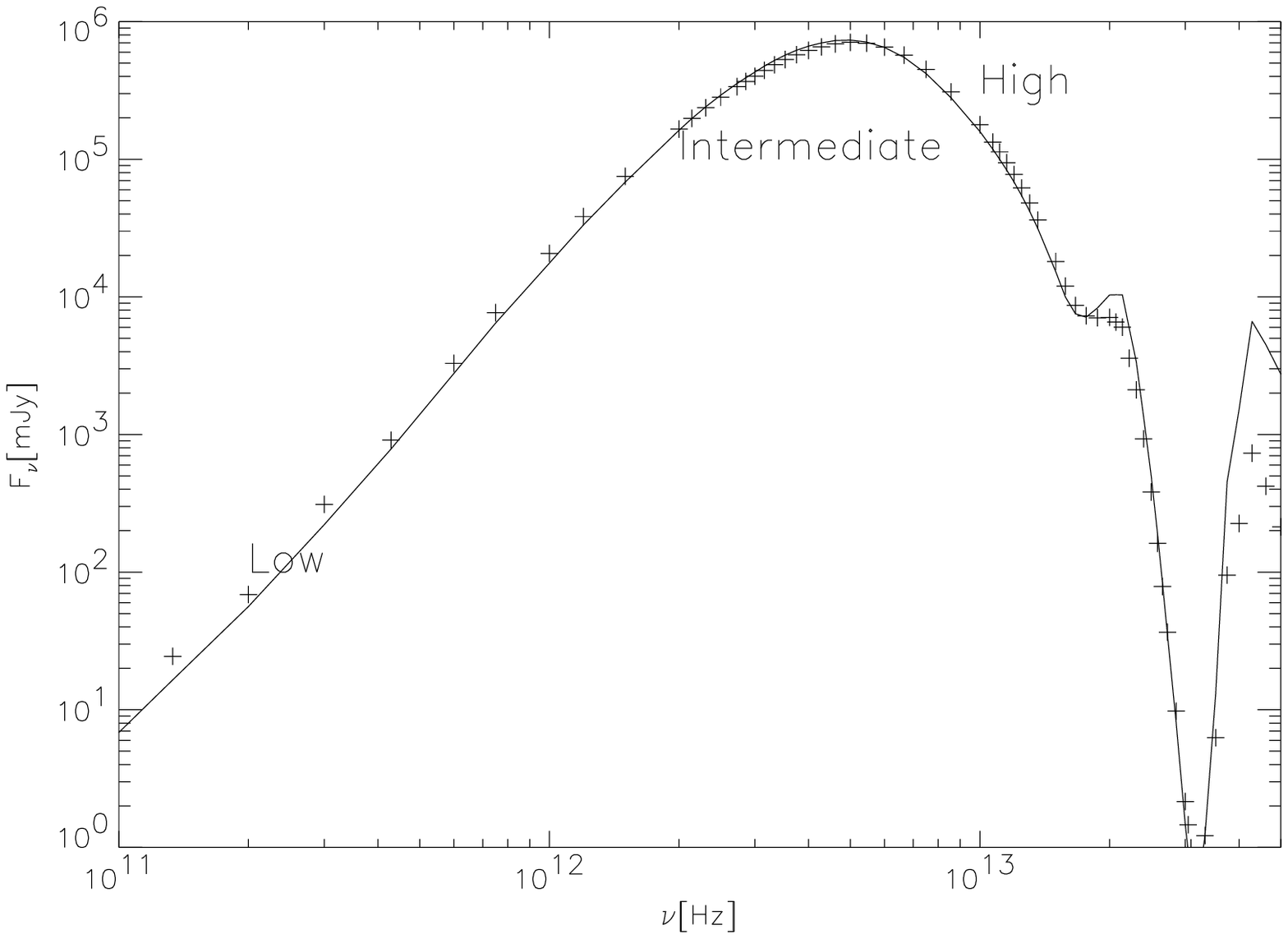,height=2.8in,width=2.8in}
\psfig{file=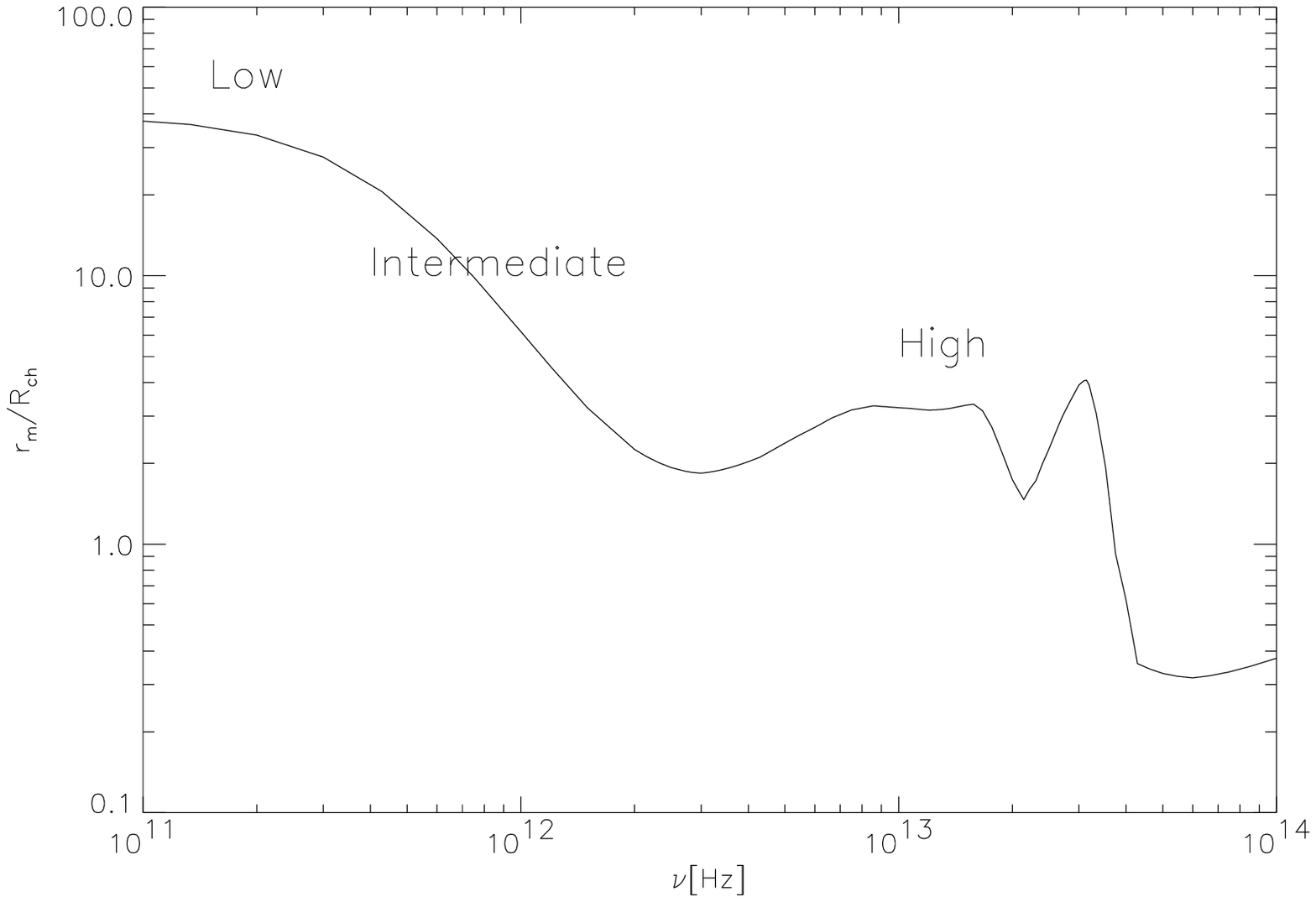,height=2.8in,width=2.8in}
{\psfig{file=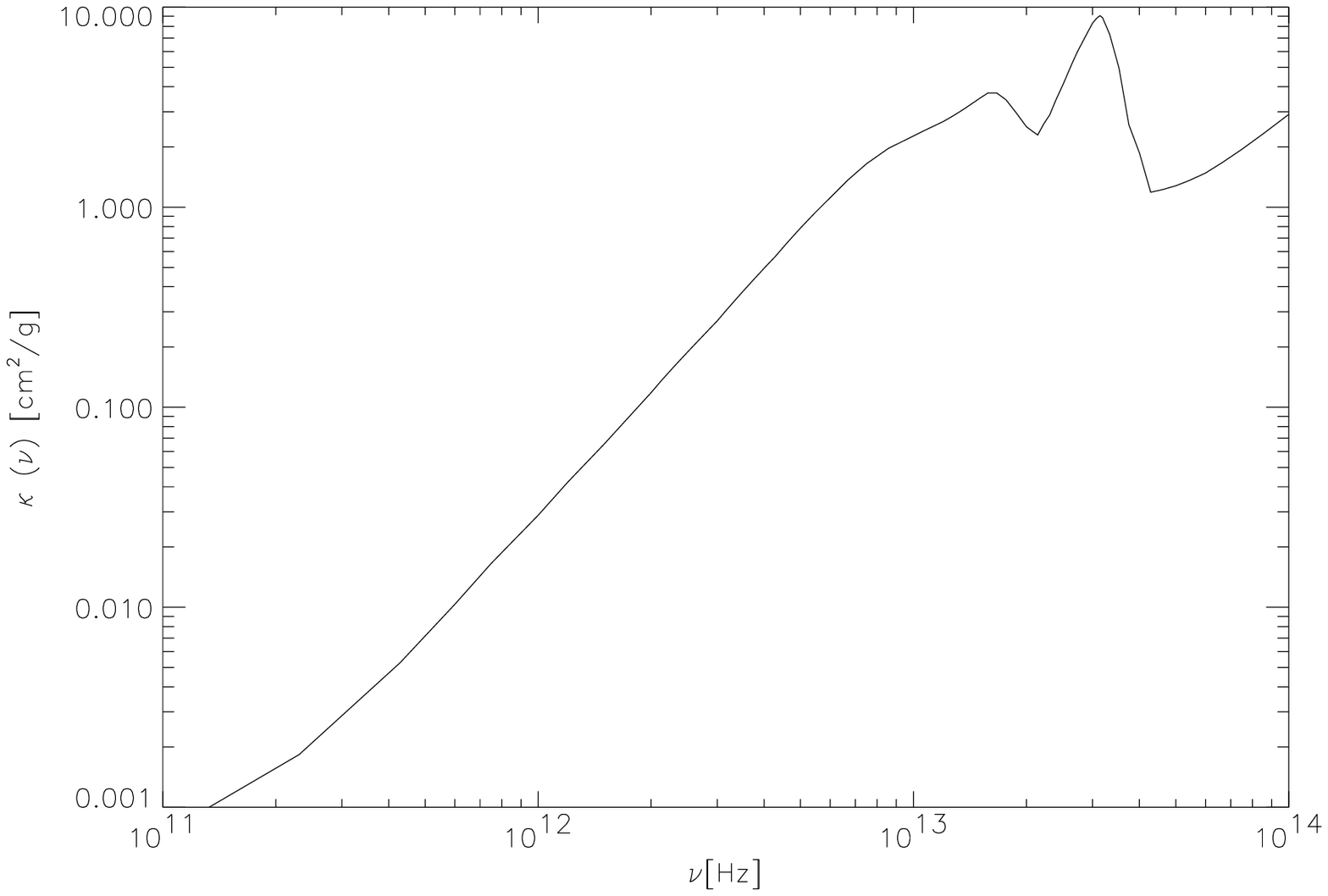,height=2.8in,width=2.8in}}}
\end{center}
\caption{(a) SED for typical high mass protostar with frequency regimes marked (solid line is DUSTY SED and crosses analytic SED) (b) Contribution function (the characteristic emission radius), in dimensionless units, $\tilde{r}_{m}$, with frequency regimes marked, (c) WD01 opacity curve.  The spectral features in the SED and opacity curve as shown in (a) and (c), e.g. the $3\times 10^{13}\rm Hz$ ($10 \micron$) absorption feature, correlate with the location in the envelope this emission is coming from, as shown in (b)}
\end{figure}

\begin{figure}[h] \begin{center}
\centerline{\psfig{file=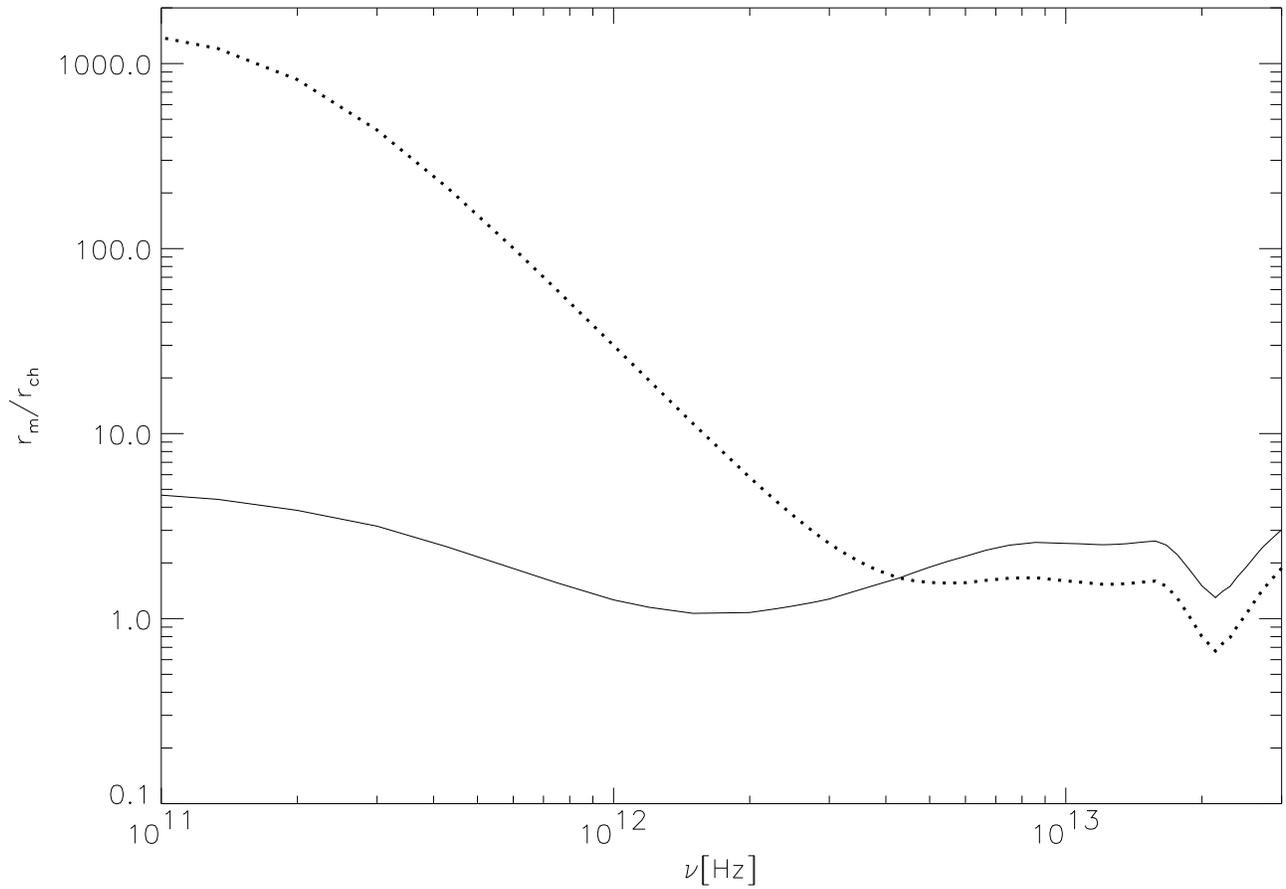}}
\end{center}
\caption{Contribution function (the characteristic emission radius), in dimensionless units, $\tilde{r}_{m}$, for a compact envelope, $\rct=4$ (solid line), and for an extended envelope, $\rct=2000$ (dotted line)}
\end{figure}

\begin{figure}[h] \begin{center}
\centerline{\psfig{file=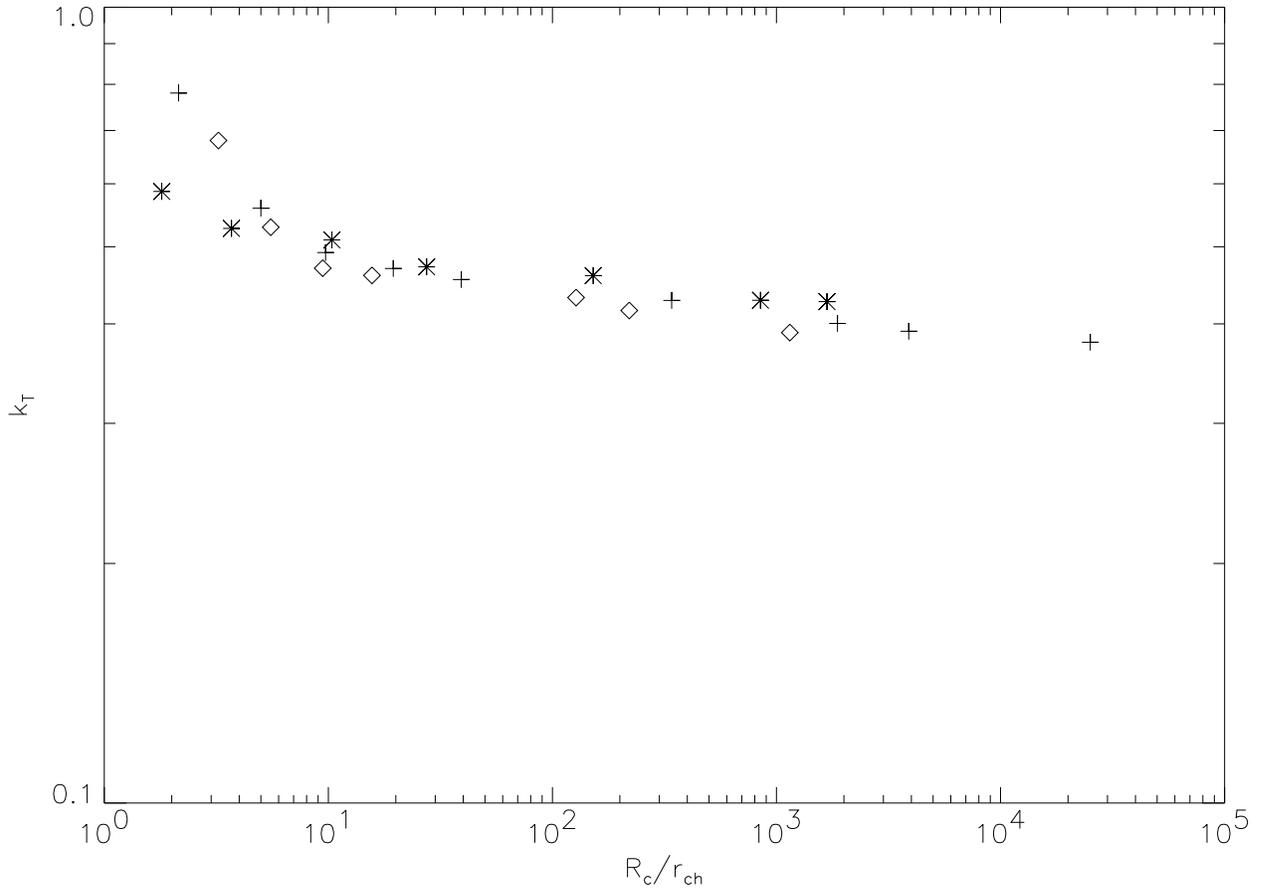}}
\end{center}
\caption{(a)Slope of temperature profile, $k_{T}$ as a function of $\rct$.  Asterisks denote $k_{\rho}=1$, crosses denote $k_{\rho}=3/2$, and diamonds denote $k_{\rho}=2$}
\end{figure}

\begin{figure}
\centerline{\psfig{file=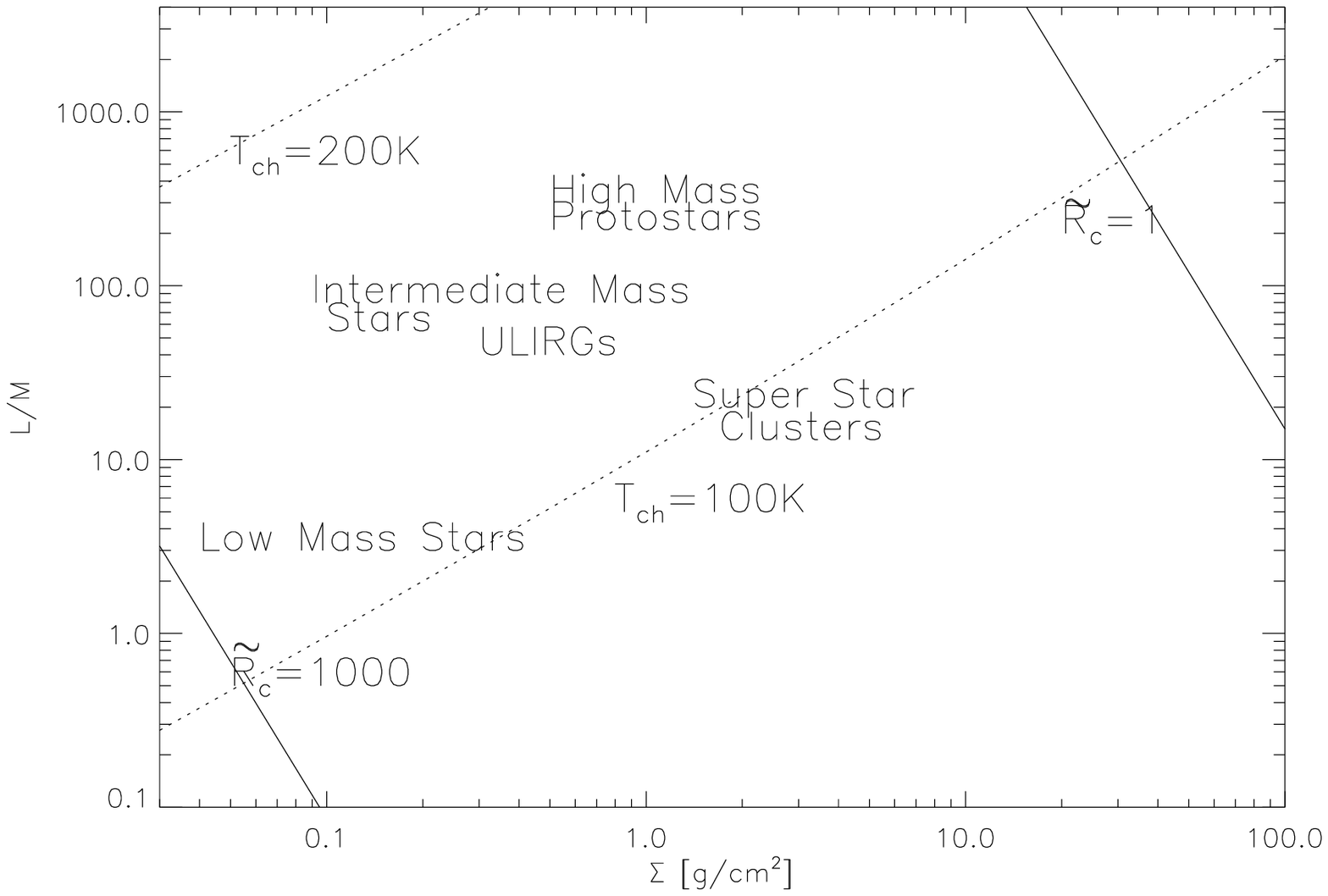}}
\caption{$L/M [L_{\odot}/M_{\odot}]$ vs $\Sigma [\rm g/cm^{2}]$ for  $k_{\rho}=2$}
\end{figure}

\begin{figure}
\centerline{\psfig{file=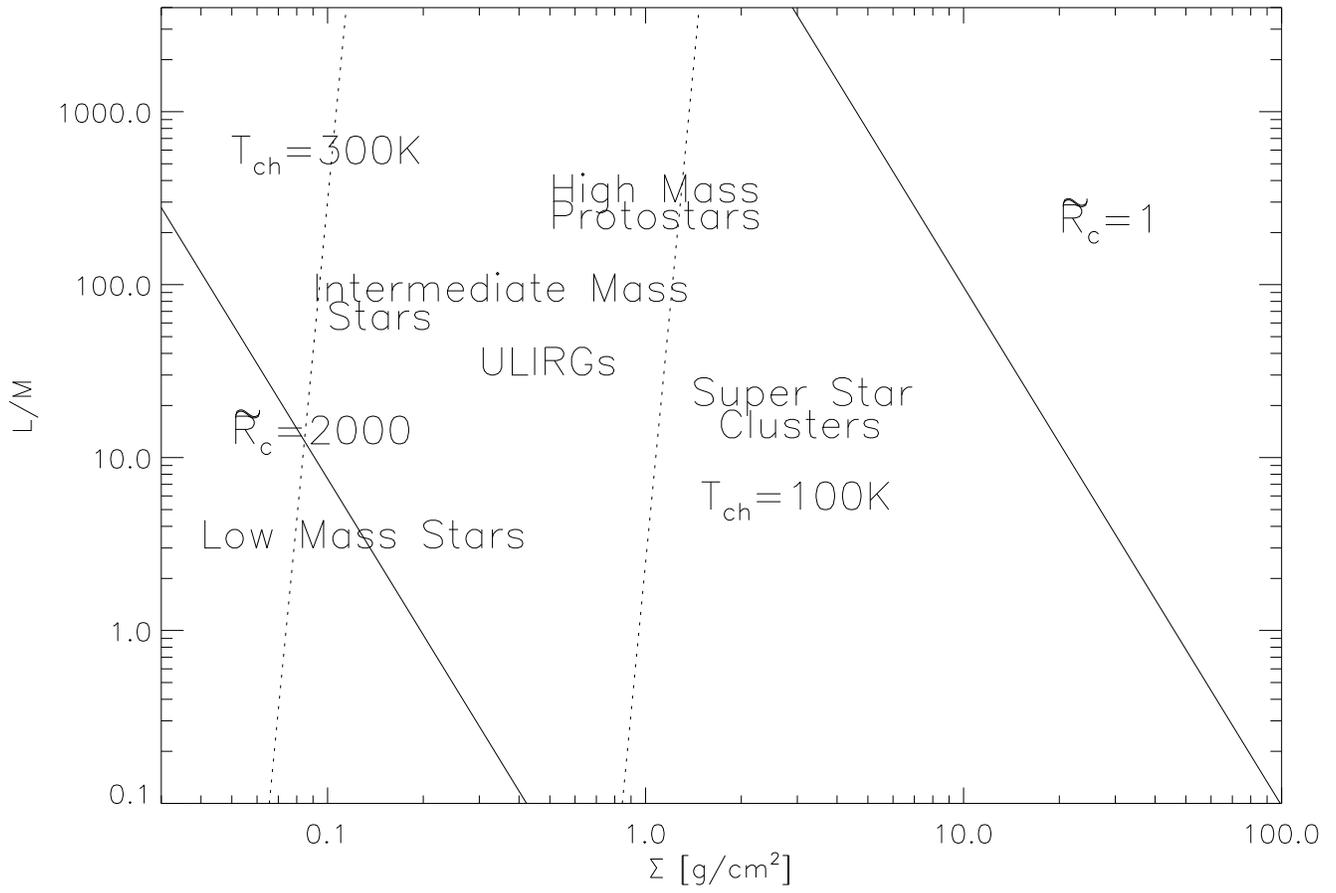}}
\caption{$L/M [L_{\odot}/M_{\odot}]$ vs $\Sigma [\rm g/cm^{2}]$ for  $k_{\rho}=1$} 
\end{figure}

\begin{figure}[h] \begin{center}
\centerline{\psfig{file=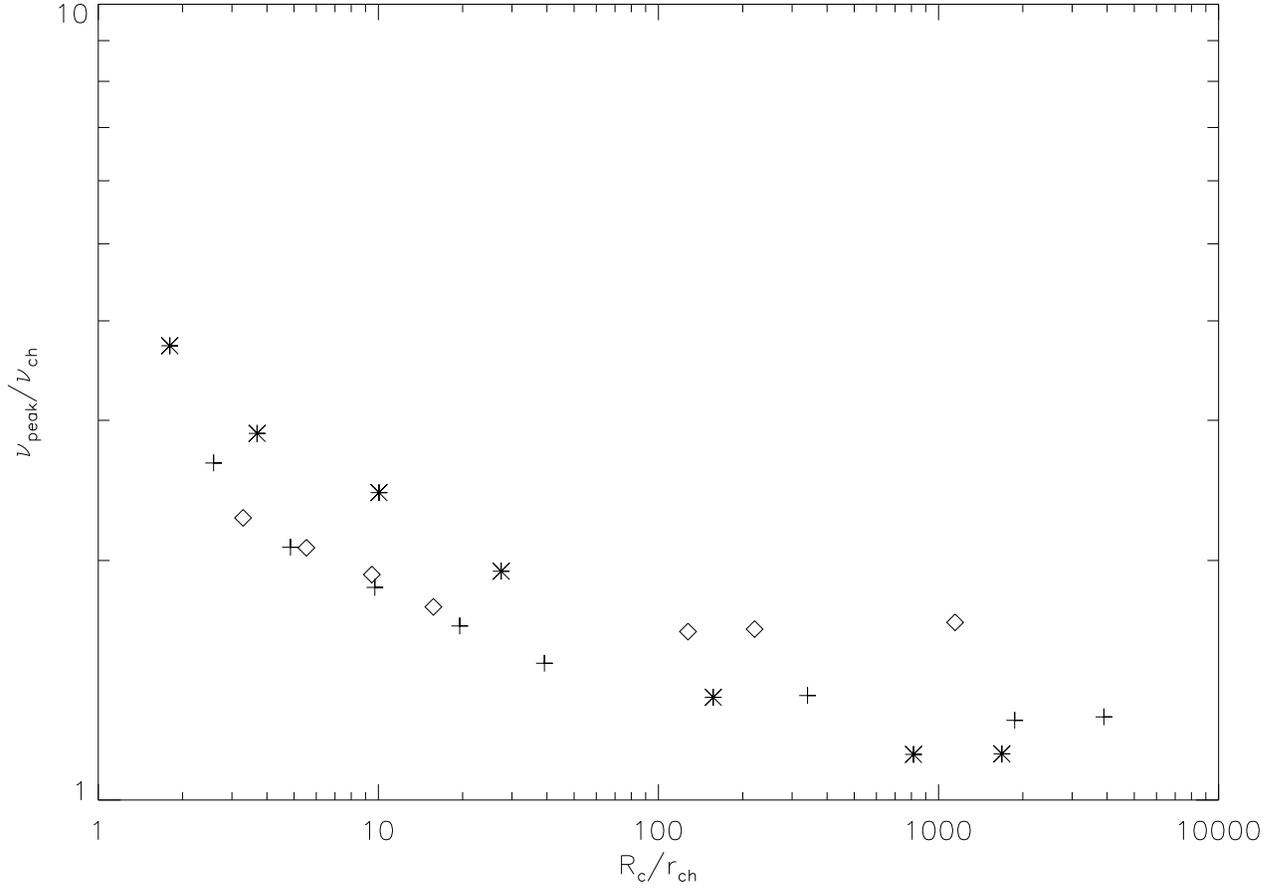}}
\end{center}
\caption{The ratio of the peak frequency to the characteristic frequency, $\nu_{peak}/\nu_{ch}$, as a function of $\rct$.  Asterisks denote $k_{\rho}=1$, crosses denote $k_{\rho}=3/2$, and diamonds denote $k_{\rho}=2$.}
\end{figure}

\begin{figure}[h] \begin{center}
\centerline{\psfig{file=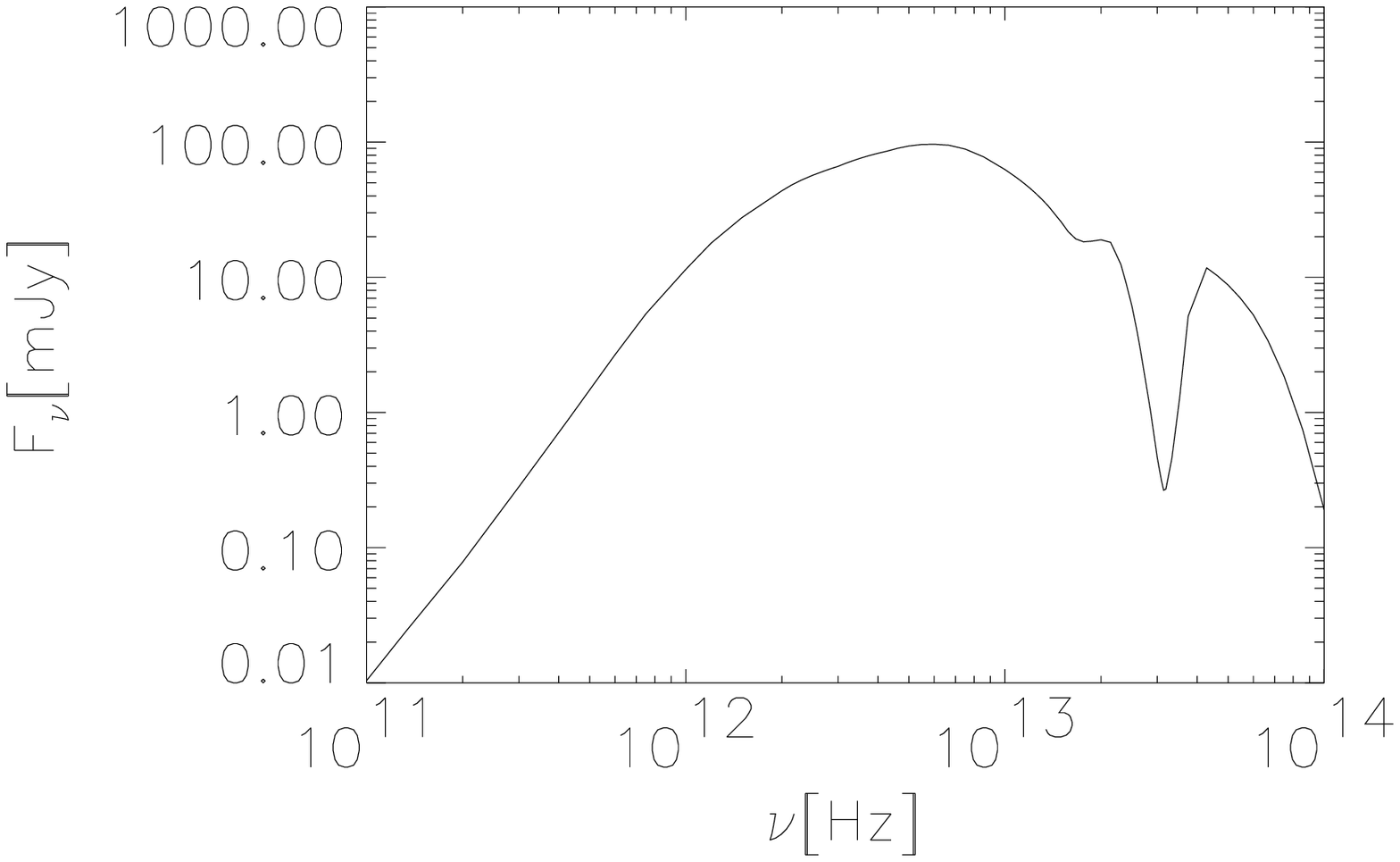,height=3.3in,width=3.3in}
\psfig{file=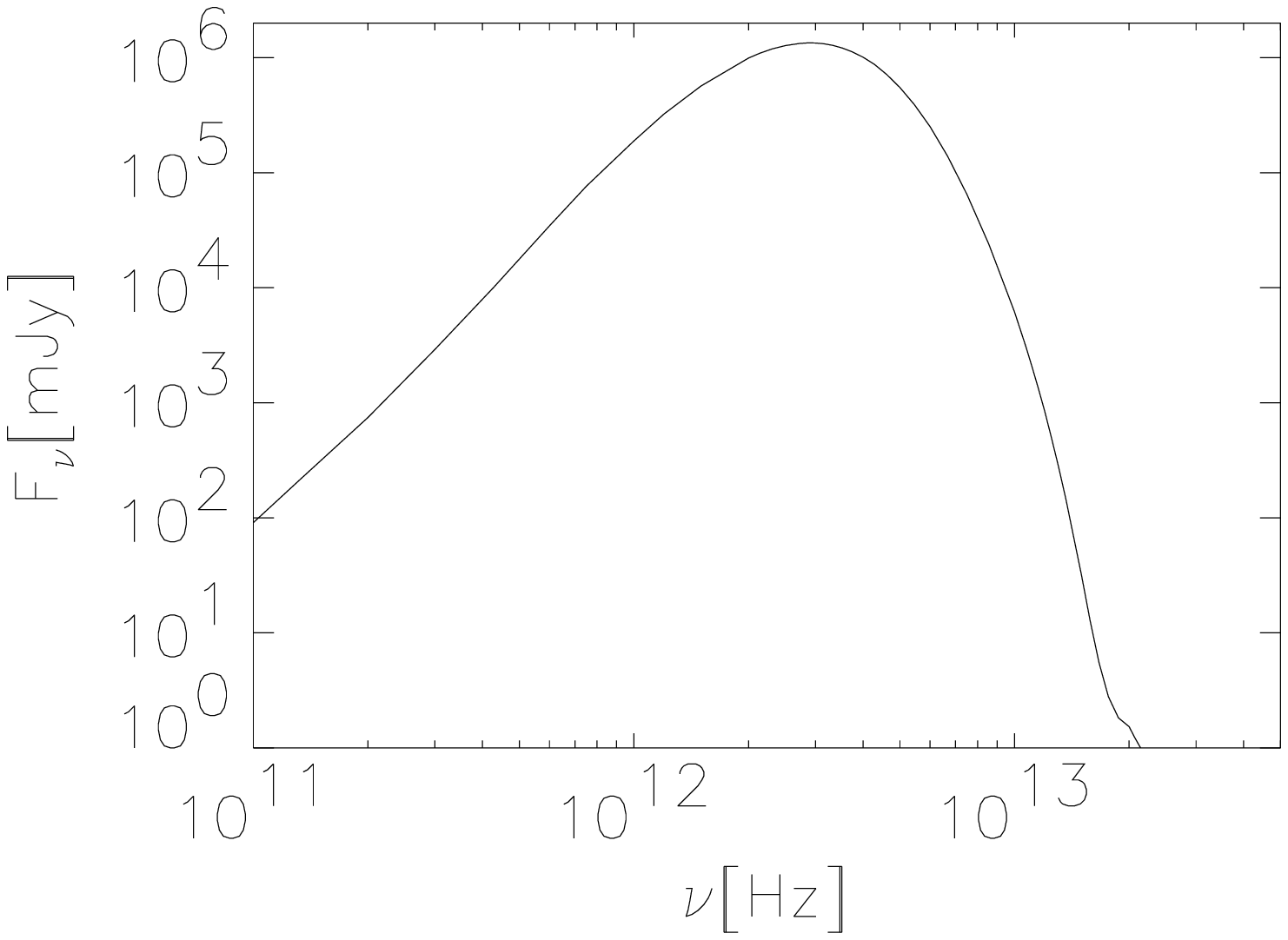,height=3.3in,width=3.3in}}
\end{center}
\caption{(a) Far-IR SED for typical low mass protostar, (b) Far-IR SED of typical ULIRG}
\end{figure}

\begin{figure}[h] \begin{center}
\centerline{\psfig{file=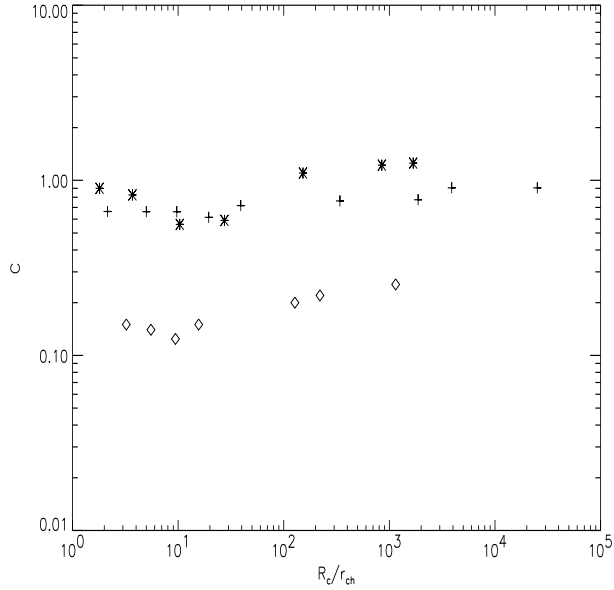,height=3.3in,width=3.3in}}
\end{center}
\caption{$C(\rct)$ for $k_{\rho}=3/2$ (crosses), $k_{\rho}=1$ (asterisks), and $k_{\rho}=2$ (diamonds)}
\end{figure}

\begin{figure}[h] \begin{center}
\centerline{\psfig{file=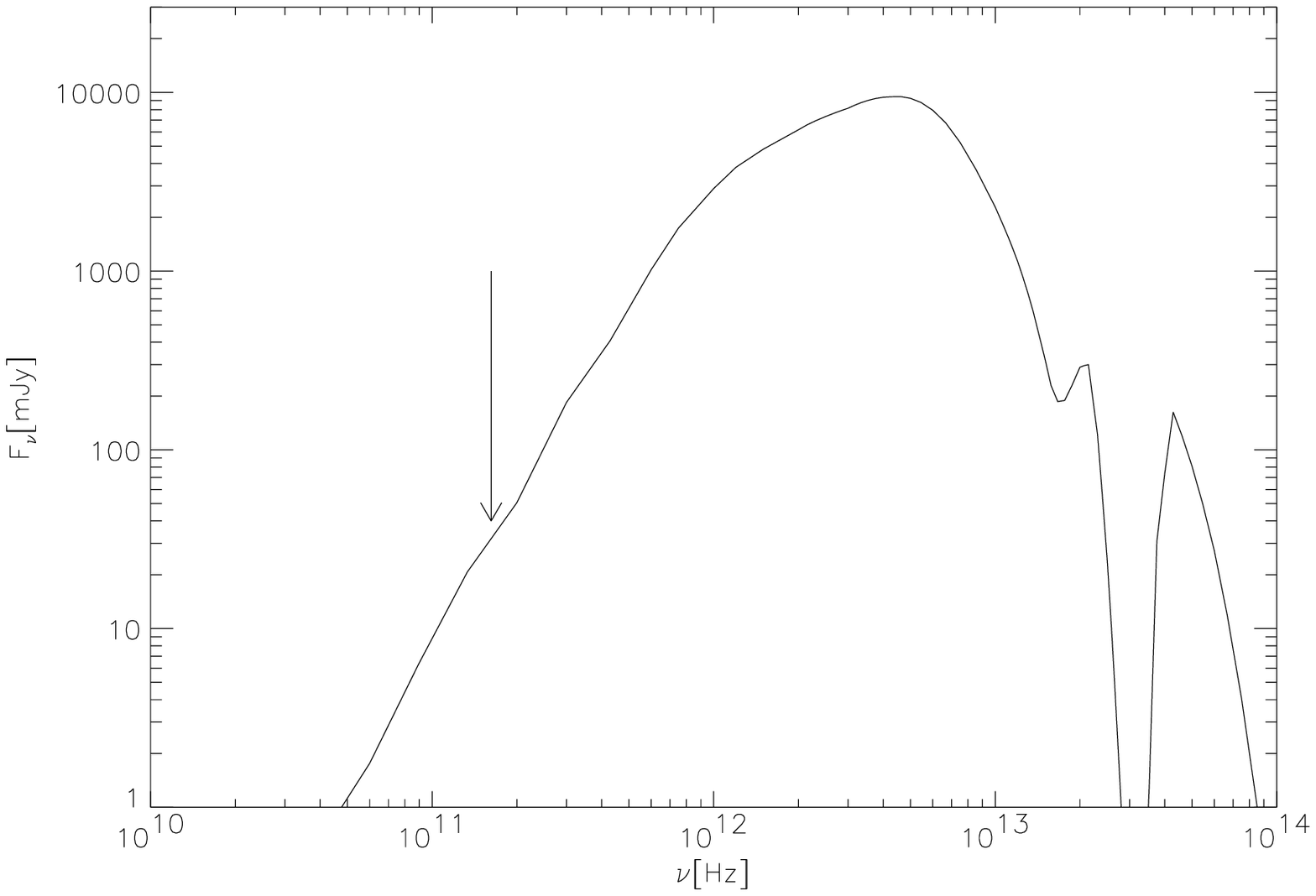,height=3.3in,width=3.3in}
\psfig{file=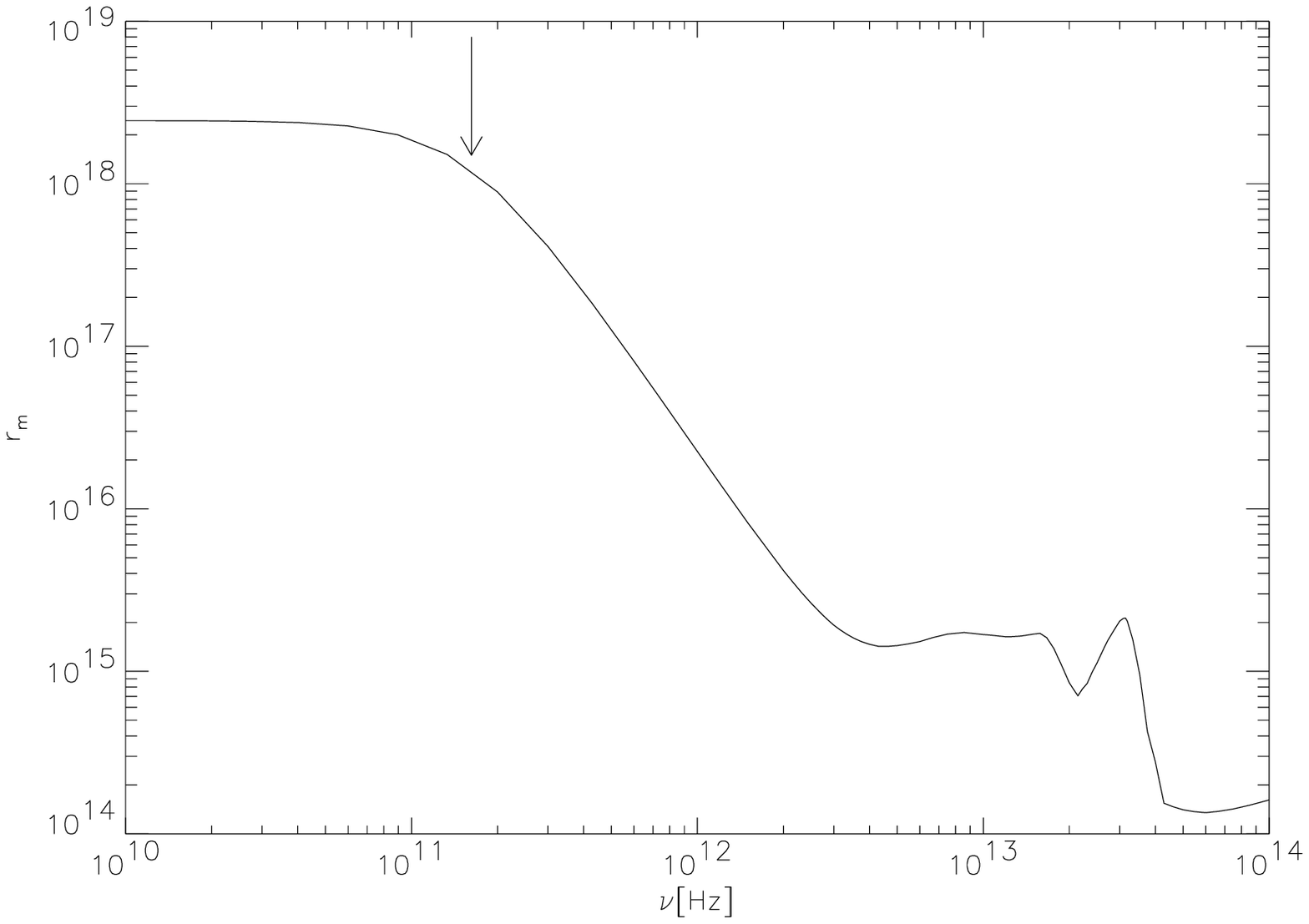,height=3.3in,width=3.3in}}
\end{center}
\caption{(a) SED for large $\rct$ envelope (extended envelope) with break frequency marked, (b) Contribution function}
\end{figure}

\begin{figure}[h] \begin{center}
\centerline{\psfig{file=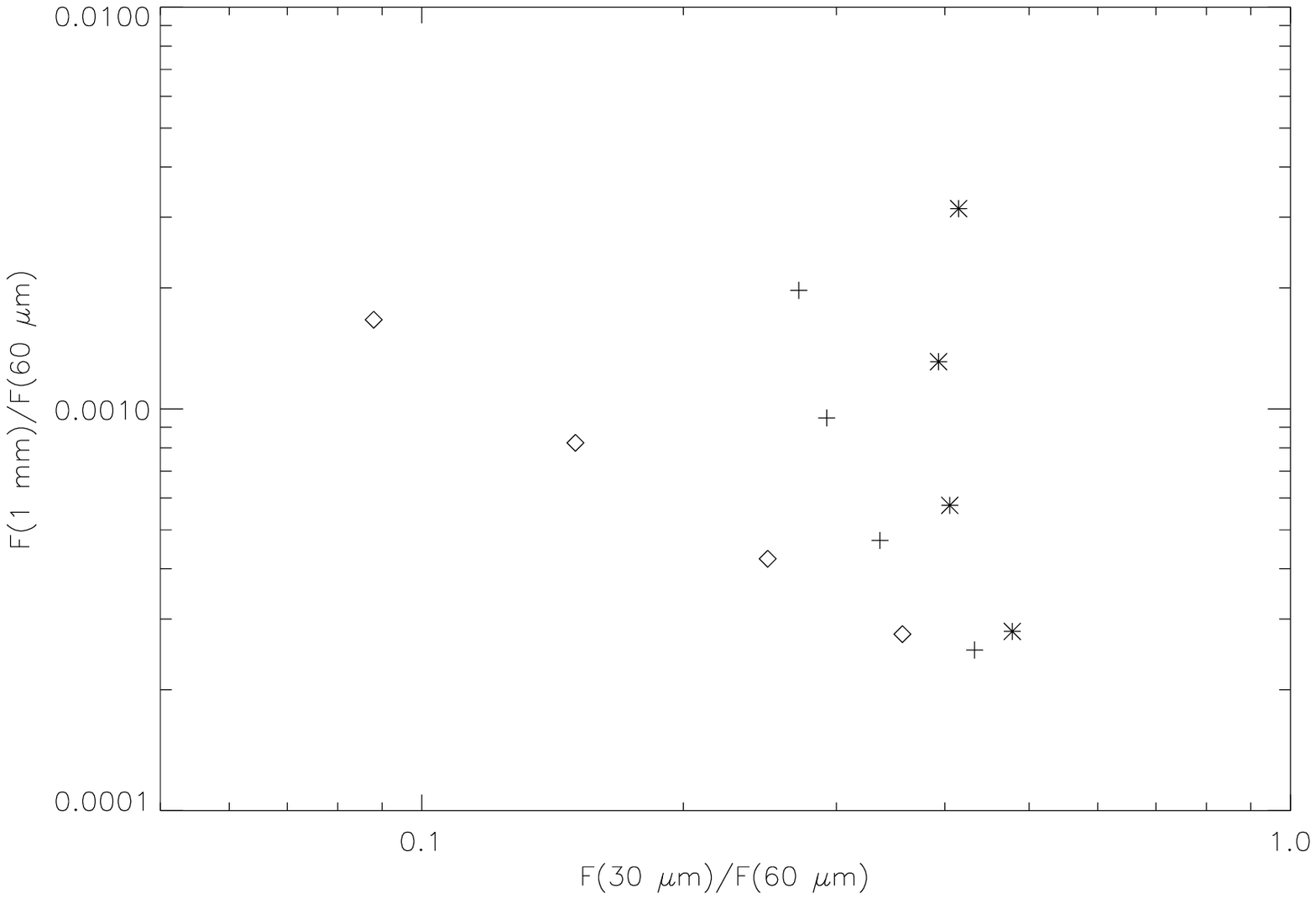,height=3.3in,width=3.3in}
\psfig{file=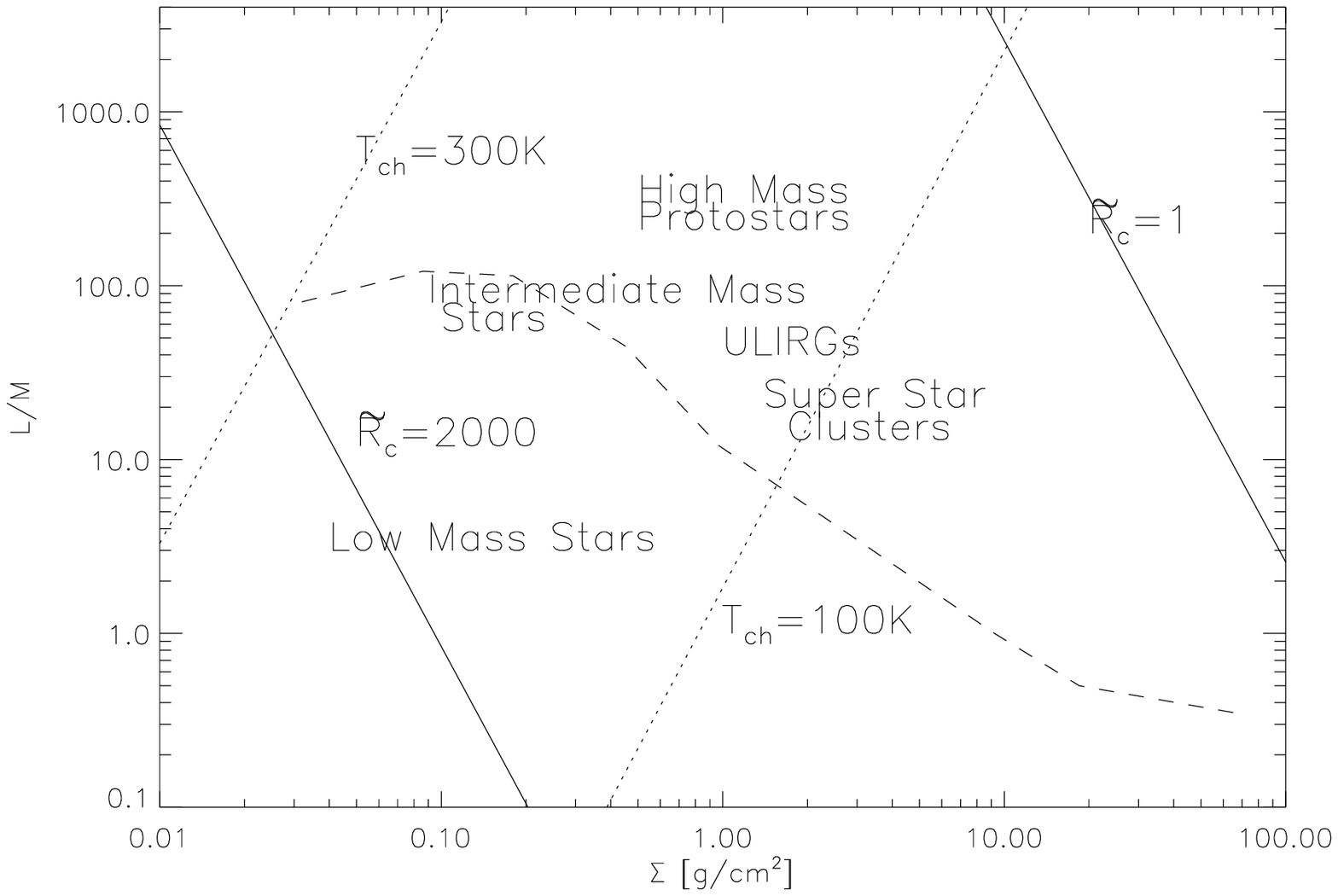,height=3.3in,width=3.3in}}
\end{center}
\caption{(a)$F(1 \rm mm)/F(60 \rm \mu m)$ vs $F(30 \rm \mu m)/F(60 \rm \mu m)$ for $k_{\rho}=2$ (diamonds), $k_{\rho}=3/2$ (crosses), $k_{\rho}=1$ (asterisks); $k_{\rho}=3/2$ curve corresponds to $\tch=210$ isotherm, (b) $L/M, \Sigma$ plot for $k_{\rho}=3/2$, dashed line demarcates region below which density profiles can be inferred from far-IR SED}
\end{figure}

\end{document}